%% file: nonlinear-bec.tex
\documentclass[11pt,a4paper]{article}
\usepackage{jheppub}
\pdfoutput=1

\usepackage{amsmath,amssymb}
\usepackage{bm}
\usepackage{graphicx}
\graphicspath{{figures/}}
\usepackage[utf8]{inputenc}
\usepackage[normalem]{ulem} 
\usepackage[shortlabels]{enumitem}
\usepackage{layouts}

\newcommand{\singleWid}{4.4475in}
\newcommand{\fullWid}{5.93in}

\newcommand{\figref}[1]{fig.~\ref{#1}}
\newcommand{\Figref}[1]{Fig.~\ref{#1}}
\newcommand{\secref}[1]{section~\ref{#1}}

\newcommand{\appref}[1]{appendix~\ref{#1}}
\newcommand{\Appref}[1]{Appendix~\ref{#1}}
\newcommand{\myref}{ref.\,}
\newcommand{\myrefs}{refs.\,}

\newcommand{\ie}{i.e.,\,\,}
\newcommand{\eg}{e.g.,\,\,}
\newcommand{\etal}{et al.}

\input{notation.tex}
\newcommand{\kdim}{\nodim{k}}

\begin{document}

\title{Nonlinear Dynamics of the Cold Atom Analog False Vacuum}

\author[a,b]{Jonathan Braden}
\author[c,d]{Matthew C.\ Johnson}
\author[b,e]{Hiranya V.\ Peiris}
\author[b]{Andrew Pontzen}
\author[f,g]{Silke Weinfurtner}

\affiliation[a]{Canadian Institute for Theoretical Astrophysics, University of Toronto, 60 St. George St, Toronto, ON, M5S 3H8, Canada}
\affiliation[b]{Department of Physics and Astronomy, University College London, Gower Street, London, WC1E 6BT, UK}
\affiliation[c]{Department of Physics and Astronomy, York University, Toronto, ON, M3J 1P3, Canada}
\affiliation[d]{Perimeter Institute for Theoretical Physics, 31 Caroline St. N, Waterloo, ON, N2L 2Y5, Canada}
\affiliation[e]{Oskar Klein Center for Cosmoparticle Physics, Department of Physics, Stockholm University, AlbaNova, Stockholm, SE-106 91, Sweden}
\affiliation[f]{School of Mathematical Sciences, University of Nottingham, University Park, Nottingham, NG7 2RD, UK}
\affiliation[g]{Centre for the Mathematics and Theoretical Physics of Quantum Non-Equilibrium Systems, University of Nottingham, Nottingham NG7 2RD, UK}

\emailAdd{jbraden@cita.utoronto.ca}
\emailAdd{mjohnson@perimeterinstitute.ca}
\emailAdd{h.peiris@ucl.ac.uk}
\emailAdd{a.pontzen@ucl.ac.uk}
\emailAdd{silke.weinfurtner@nottingham.ac.uk}


\abstract{
We investigate the nonlinear dynamics of cold atom systems that can in principle serve as quantum simulators of false vacuum decay. The analog false vacuum manifests as a metastable vacuum state for the relative phase in a two-species Bose-Einstein condensate (BEC), induced by a driven periodic coupling between the two species. In the appropriate low energy limit, the evolution of the relative phase is approximately governed by a relativistic wave equation exhibiting true and false vacuum configurations. In previous work, a linear stability analysis identified exponentially growing short-wavelength modes driven by the time-dependent coupling. These modes threaten to destabilize the analog false vacuum. Here, we employ numerical simulations of the coupled Gross-Pitaevski equations (GPEs) to determine the non-linear evolution of these linearly unstable modes. We find that unless a physical mechanism modifies the GPE on short length scales, the analog false vacuum is indeed destabilized.  We briefly discuss various physically expected corrections to the GPEs that may act to remove the exponentially unstable modes.  To investigate the resulting dynamics in cases where such a removal mechanism exists, we implement a hard UV cutoff that excludes the unstable modes as a simple model for these corrections.  We use this to study the range of phenomena arising from such a system. In particular, we show that by modulating the strength of the time-dependent coupling, it is possible to observe the crossover between a second and first order phase transition out of the false vacuum.
}

\maketitle

\section{Introduction}\label{sec:intro}
False vacuum decay, the quantum first-order phase transition out of a metastable ``false vacuum'' state, plays an important role in current models of the early Universe. 
This includes theories of false vacuum eternal inflation (see \eg \myref~\cite{Guth:2007ng}), symmetry breaking phase transitions in GUT theories~\cite{Guth:1981uk}, and perhaps even the Standard Model of particle physics~\cite{Ellis:2009tp}.
The dynamics of these transitions has implications for anthropic solutions to the cosmological constant problem, the observability of false vacuum eternal inflation in the multiverse~\cite{Aguirre:2007an,Feeney:2010dd,Feeney:2010jj}, the future fate of the Higgs vacuum~\cite{Ellis:2009tp}, and primordial gravitational wave signals~\cite{Kosowsky:1991ua} within the band of future experiments such as the Laser Interferometer Space Antenna (LISA) (see \eg \myref~\cite{Axen:2018zvb}). 

Further, as an intrinsically quantum mechanical phenomenon (\ie requiring $\hbar \neq 0$), false vacuum decay touches on fundamental issues in quantum field theory, including the notions of measurement and the emergence of classicality.
Therefore, a complete computational and conceptual framework to understand false vacuum decay has implications both for our understanding of the cosmos and for the foundations of quantum field theory.

In the standard view, false vacuum decay is essentially a field theory version of tunneling through a barrier familiar from quantum mechanics~\cite{Coleman:1977py,Callan:1977pt,Coleman:1980aw}.
However, an important distinction between quantum mechanics and quantum field theory is that the former involves only a single degree of freedom, while the latter involves interactions between infinitely many degrees of freedom~\cite{Banks:1973ps}.
This has important consequences, both conceptually and computationally.
In the case of quantum mechanics, it is straightforward to evolve the wavefunction using the Schr\"{o}dinger equation, thus obtaining a complete numerical solution to the tunneling problem.
However, an analogous approach is infeasible in quantum field theory due to the exponential complexity of the state space.
In the standard formalism~\cite{Coleman:1977py,Callan:1977pt}, false vacuum decay is reformulated as a quantum mechanics problem, resulting in a drastic dimensional reduction of the underlying phase space.
Bubble nucleation is then interpreted as a quantum tunneling event, with no real-time description available.
We recently presented an alternative description of vacuum decay based on a careful study of dynamics in the full field space, considering the cooperative dynamics of many field degrees of freedom~\cite{Braden:2018tky}.
Using this approach, we were able to reproduce the expected decay rates of the standard formalism, while also giving a real-time description of the decay process that does not rely on quanutm tunneling.
Given the differing approximations and subsequent interpretations of these two approaches, it would be greatly informative to study a physical system that undergoes false vacuum decay, where Nature automatically includes all quantum effects.

One proposal to build such a system uses a two-component dilute gas cold atom Bose-Einstein condensate (BEC) to simulate a relativistic scalar field with a false vacuum potential~\cite{Fialko:2014xba,Fialko:2016ggg}.
In this proposal, the false vacua are generated from unstable local maxima by periodically modulating the direct conversion between the two components.
From the field theory perspective, this amounts to modulating the overall amplitude of the potential, which can lead to a stabilization of the long-wavelength modes as in the Kapitza pendulum.
However, as recently shown by Braden \etal~\cite{Braden:2017add}, this simultaneously destabilizes shorter wavelength modes, whose exponential growth is expected to lead to a breakdown of the analogy with the relativistic scalar field system, and a classical destabilization of the false vacuum.
Therefore, the feasibility of such an experiment rests on a physical mechanism to remove these dynamically unstable short-wavelength modes, leaving only the long-wavelength degrees of freedom for which the false vacuum description is valid. Another recent work explored the role of impurities (vortices) in the condensate in seeding vacuum decay~\cite{Billam:2018pvp}.

In this paper, we explore the viability of the analog false vacuum in detail through the use of nonlinear simulations of cold atom BECs.
Our primary focus is on the viability of generating a metastable state within the BEC, rather than providing a detailed dictionary between the BEC system and a relativistic scalar field.
First, we verify the destabilization of the condensate by the parametrically excited short-wavelength instabilities, confirming the results obtained in the linear analysis of \myref~\cite{Braden:2018tky}.
We also briefly explore the subsequent nonlinear dynamics.
In this case, not only is the false vacuum interpretation of the experiment disrupted, but the entire relativistic scalar field analogy breaks down.
In order to restore the false vacuum interpretation, corrections to the GPE description must appear at wavenumbers around the Floquet band to either damp out or otherwise remove these exponentially unstable modes.
Simultaneously, these corrections should be negligible for the longer wavelength modes needed to form the bubbles.
In order to be viable, such corrections must be motivated by real physical effects, rather than numerical tricks such as a convenient choice of grid spacing.
We briefly outline some physically plausible modifications to the short-wavelength dynamics of the condensate.
Since modifications to the GPE at short-wavelengths will generically leak into the dynamics of modes required to form bubbles, the impact of these modifications must be accounted for when making quantitative predictions.
A detailed study of these subtleties is beyond the scope of this work and will be explored in future publications.
Regardless of the scalar field reinterpretation, the BEC system still represents a metastable state that decays as a result of quantum effects (if the system is safe from the unstable Floquet modes).
Therefore, even in the absence of a direct link to early Universe models, such systems can be used to investigate various fundamental issues in quantum field theory.

The remainder of this paper is organized as follows. First, \secref{sec:bec-review} presents the theoretical description of dilute gas cold atom Bose-Einstein condensates, modeled by the Gross-Pitaevskii equation (GPE).  Next, \secref{sec:effective-scalar} briefly reviews the connection between the GPE and the dynamics of a relativistic scalar field, in particular how to generate a false vacuum potential minimum. 
Our main results are contained in \secref{sec:floquet-modes} and \secref{sec:nonlinear-scalar}.
In~\secref{sec:floquet-modes}, we explore the manifestation of Floquet instabilities in the full nonlinear condensate dynamics, verifying that they lead to a breakdown of the effective relativistic field theory description of the condensate fluctuations.
We briefly comment on some physical effects that could tame these instabilities and restore the analog false vacuum interpretation.
Meanwhile, \secref{sec:nonlinear-scalar} explores the range of nonlinear field theory dynamics that can arise once the Floquet modes have been exorcised, showing how we can slowly deform a second-order phase transition into a first-order phase transition through the tuning of an experimental parameter.
Finally, we conclude in~\secref{sec:conclusions}.
Some more technical aspects of the analysis are presented in a set of appendices.
In \appref{app:trunc-wigner} we briefly outline our semi-classical lattice approach---the truncated Wigner approximation---used for numerical simulations.
\Appref{app:dim-var} introduces our dimensionless units and equations of motion, and provides a dictionary between our choices and the previous literature.
Some further evidence that we are seeing the Floquet instability and details of the deformed linear dispersion relationship induced by finite-differencing stencils are given in~\appref{app:effective-k}.
Finally, tests of our numerical code are presented in~\appref{app:convergence}, including direct convergence tests and a demonstration that relevant Noether charges are conserved.

\section{2-Component Cold Atom Bose-Einstein Condensates in the Dilute Gas Limit}\label{sec:bec-review}
The analog false vacuum system under investigation is a 2-component BEC. There are currently two different configurations that can in principle be utilized as false vacuum decay simulators. The first one is a single-species BEC in a double-well potential in the so-called tight-binding regime, exhibiting symmetric and anti-symmetric single-particle states~\cite{Albiez:2005,Bucker:2011}.  False vacuum decay requires a modulation of the tunnel-coupling via the trapping parameters. At present, this configuration is capable of mimicking false vacuum decay in $1$+$1$-dimensions.  To avoid dimensional restrictions altogether, we focus on a different configuration: a two species BEC of atoms with mass $m$, \eg $^{41}$K or $^{87}$Rb atoms in two different hyperfine states with a modulated linear coupling between the two species.
In this paper we will consider the case of a condensate confined to a ring-trap, which we model as a $1$+$1$-dimensional condensate.
The analogy between the two species system and false vacuum decay has been derived in detail in \myref~\cite{Braden:2017add}, although the derivation applies to the double-well setup as well.
In the following, we will summarize the key results of \myref~\cite{Braden:2017add}, which motivate the nonlinear numerical simulations presented in the subsequent sections.  

We consider an ultra-cold highly diluted 2-component system consisting of two single-particle states ($\vert 1 \rangle$ and $\vert 2\rangle$) of mass $m_{1}=m_{2}=m$. Within this setup the atom-atom interactions can be approximated by $S$-wave contact potentials: $g_{11}$, $g_{22}$ and $g_{12}=g_{21}$. Atoms in different hyperfine states have angular momentum and hence a slightly different energy. It is possible to drive transitions between two hyperfine states with a given transition rate $\nu$ through the utilization of external fields, \eg a radio-frequency field. When the BEC forms, the collective long-wavelength excitations of the condensed atoms (\ie the condensate) acquire a vacuum expectation value (vev), which we denote by the complex numbers $\wf_i$. The Hamiltonian density for the two interacting condensates is given by 
\begin{equation}
  \mathcal{H} = \frac{\hbar^2}{2m}\abs{\nabla\wf_i}^2 + \frac{g_{ij}}{2}\abs{\wf_i}^2\abs{\wf_j}^2 - \frac{\nuT}{2}\sigma_{ij}^{\bf \rm x}\left(\wf_i\wf_j^*+\wf_i^*\wf_j\right) \, .
\end{equation}
Here we are dropping the effects of the external confining potentials $V_{\rm ext,i}$ on the dynamics, and we have defined
\begin{equation}
  \sigma^{\rm \bf x} = \left(\begin{array}{cc}0 & 1 \\ 1 & 0 \end{array}\right) \, .
\end{equation}
The combined evolution of the 2-component system is given by
\begin{subequations}\label{eqn:gpe-cartesian}
  \begin{align}
    i\hbar\dot{\wf}_1 &= -\frac{\hbar^2\nabla^2}{2m}\wf_1 + \gS\abs{\wf_1}^2\wf_1 + \gC\abs{\wf_2}^2\wf_1 - \nuT\wf_2 \, ,\\
    i\hbar\dot{\wf}_2 &= -\frac{\hbar^2\nabla^2}{2m}\wf_2 + \gS\abs{\wf_2}^2\psi_2 + \gC\abs{\wf_1}^2\wf_2 - \nuT\wf_1 \, ,
  \end{align}
\end{subequations}
where we have assumed the inter-atomic scattering strengths are of the form $g_{11}=g_{22}=\gS$ and defined $g_{12}=\gC$.
The effects of quantum fluctuations are then approximated by generating realizations of complex Gaussian random fields, and superimposing these on the appropriate background homogeneous values of $\wf_i$.
These serve as the initial conditions for subsequent evolution of the coupled GPEs.
The entire procedure is known as the truncated Wigner approximation.
Further details about the truncated Wigner approximation, as well as our numerical approach to solve the coupled GPEs and generate initial quantum fluctuations, are given in~\appref{app:trunc-wigner}.
For the remainder of this work, we consider only the limit $g_{11}=g_{22}=\gS$ and $\gC = 0$, which we refer to as a symmetric theory.
To generate an effective false vacuum potential, we further consider a periodically modulated coupling,
\begin{equation}
  \nuT(t) = \nuZero + \nuAmp\hbar\nuFreq\cos(\nuFreq t) \, .
\end{equation}

In order to connect with the dynamics of relativistic scalar fields, it is convenient to consider an alternative set of Hermitian canonical coordinates, the condensate densities $\dens_i$ and phases $\phase_i$, such that
\begin{equation}
  \wf_i = \sqrt{\dens_i}e^{i\phase_i} \; \mbox{and} \; \wf_i^* = \sqrt{\dens_i}e^{-i\phase_i}.
\end{equation}
In terms of these, the coupled GPEs~\eqref{eqn:gpe-cartesian} become
\begin{subequations}\label{eqn:gpe-dens-phase}
  \begin{align}
    \hbar\dot{\dens}_i &= -\frac{\hbar^2}{m_i}\left(\dens_i\nabla^2\phase_i + \nabla\dens_i\cdot\nabla\phase_i\right) - 2\nu\sum_{i\neq j}\sqrt{\dens_i\dens_j}\sin(\phase_j-\phase_i) , \\
    \hbar\dot{\phase}_i &= \frac{\hbar^2}{2m_i}\left(\frac{\nabla^2\dens_i}{2\dens_i} - \frac{(\nabla\dens_i)^2}{4\dens_i^2} - (\nabla\phase_i)^2\right) - g_{ij}\dens_j + \nuT\sum_{i\neq j}\sqrt{\frac{\dens_j}{\dens_i}}\cos(\phase_j-\phase_i) \, . \label{eqn:phase-ev}
  \end{align}
\end{subequations}

To map the coupled 2-component BEC onto the relativistic false vacuum decay~\cite{Braden:2017add}, it is natural to instead use the total density and phase, and the relative density and phase. For notational convenience we will denote these by
\begin{subequations}
  \begin{align}
  \phaseTot &\equiv \frac{\phase_1+\phase_2}{2}, \qquad
  \densTot \equiv \dens_1+\dens_2, \qquad \mathrm{and} \\
  \phaseRel &\equiv \phase_2-\phase_1, \qquad
  \densRel \equiv \frac{\dens_2-\dens_1}{2} \, ,
  \end{align}
\end{subequations}
respectively.
We will refer to the pairs of canonically conjugate variables $(\phaseTot,\densTot)$ and $(\phaseRel,\densRel)$ as the total and relative phonons, respectively.

\section{Effective Relativisitic Scalar Field Dynamics}\label{sec:effective-scalar}
We now briefly review the effective relativistic scalar field theory lurking within the dynamics of the cold atom BECs.
For a detailed path integral derivation and additional details, the interested reader can refer to \myrefs~\cite{Braden:2017add,Fialko:2014xba,Fialko:2016ggg}.
Analogous results can also be obtained by working directly with the equations of motion.
In order to streamline the presentation, here we provide the key results needed to interpret the remainder of the paper.  
Throughout, we assume that the GPE description of the previous section holds for all relevant modes of the condensates.

To relate the BEC dynamics to that of a relativistic scalar field, we first consider the evolution of a purely homogeneous background field configuration.
We denote solutions in the homogeneous approximation by an overbar~$\bar{\cdot}$.
Potential vacua are identified by looking for eigenstates of the Hamiltonian, with the eigenvalue ($\mu_{\rm bg}$) playing the role of the chemical potential and sets $\bg{\phaseTot} = -\frac{\mu_{\rm bg}}{\hbar} t$.\footnote{Including fluctuations changes the time evolution of the spatial average of the total phase $\langle\phaseTot\rangle_{\rm V}$ compared to the evolution in the homogeneous background, so that $\frac{{\rm d}}{{\rm d}t}\langle\phaseTot\rangle_{\rm V} \neq \frac{{\rm d}}{{\rm d}t}\bg{\phaseTot} = -\frac{\mu_{\rm bg}}{\hbar}$. This can be interpreted as a correction to the chemical potential from the fluctuations.  We verified that this effect appears in our nonlinear simulations.}
For $g_{11}=g_{22}$, we find stationary solutions when $\cos\bg{\phaseRel} = \pm 1$, with the $\cos\bg{\phaseRel} = -1 $ state having the higher energy and thus representing a potential false vacuum.
We next consider fluctuations around these stationary solutions to the homogeneous background equations:
\begin{subequations}
  \begin{align}
    \densTot(x,t) &= \bg{\densTot} + \pert{\densTot}(x,t) \\
    \phaseTot(x,t) &= \bg{\phaseTot}(t) + \pert{\phaseTot}(x,t) \\
    \densRel(x,t) &= \bg{\densRel} + \pert{\densRel}(x,t) \\
    \phaseRel(x,t) &= \bg{\phaseRel} + \pert{\phaseRel}(x,t)
  \end{align}
\end{subequations}
where an overbar $\bg{\cdot}$ indicates a solution to the coupled GPEs in the homogeneous limit.\footnote{Because of the nonlinear relationship between the Cartesian condensate variables ($\wf_i$) and the density ($\dens_i$) and phase variables ($\phase_i$), the volume averaged density and the background density differ by a UV divergent quantity $\langle\densTot\rangle_{\rm V} = \sum_i\abs{\bg{\wf}_i}^2 + \langle\abs{\pert{\wf}_i}^2\rangle_{\rm V}  \neq \sum_i \abs{\bg{\wf}_i}^2$.  The equality holds only when the fields are exactly spatially homogeneous.}
The density fluctuations are then integrated out, and only the modes well below the healing length are considered.  The result is a (nonlinear) effective Lagrangian for the phases that has the form of a pair of relativistic scalar fields.

As explicitly laid out in Braden \etal~\cite{Braden:2017add}, a number of conditions must hold for an effective relativistic scalar field theory for the fluctuations to emerge from the procedure outlined above.
These include:
\begin{enumerate}
  \item the evolution of the coupled condensates are given by the GPE;
  \item the background dynamics of the condensate are well approximated as homogeneous;
  \item the local fluctuations in the particle densities are small, and can be treated quadratically in the action;
  \item the short-time dynamics of the fluctuation modes are simple and can be integrated out to obtain an effective time-averaged description; and
  \item the relative and total phonons can be treated as effectively decoupled, leading to a truncation rather than integration of the total phase fluctuations.
\end{enumerate}
The first four of these are generic requirements, independent of the particular couplings in the GPE, while the fifth assumption (the decoupling between total and relative phonons) is specific to fluctuations in theories with $g_{11}=g_{22}$ expanded around a background state of equal densities in the two condensates.
This latter condition holds for the stationary solutions with $\cos\bg{\phaseRel} = \pm 1$.
In more general cases, the background condensates can be made spatially inhomogeneous. By modifying the kinetic term in the resulting effective Lagrangian, under certain conditions this leads to an interpretation of a scalar field evolving in a classical background (such as a gravitational field).

\begin{figure}
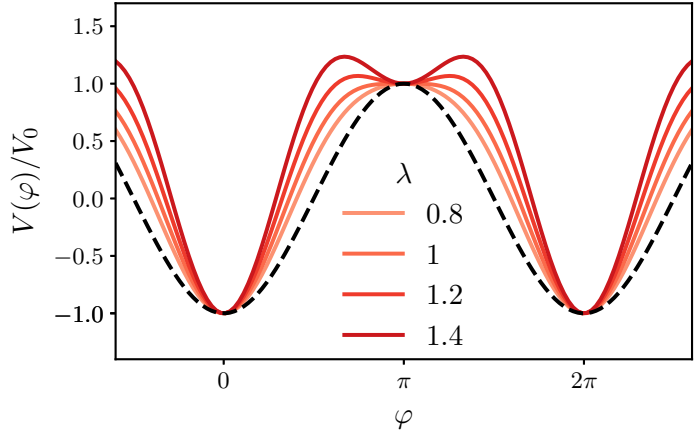
\centering
  \includegraphics[width=\singleWid]{{{effective-potentials}}}
  \caption{Time-averaged effective potential for the relative phase variables for several choices of the parameter $\lPar = \nuAmp\sqrt{\frac{2\gS \rhoBG}{\nuZero}}$.  For reference, the dashed black line is the sine-Gordon potential when $\lPar = 0$.  We see that increasing $\lPar$ transforms the local maxima of the potential into local minima.}
  \label{fig:effective-pot}
\end{figure}

Under these assumptions, the relative and total phases of the condensates behave approximately as decoupled relativistic scalar fields.
The relative phase obeys
\begin{equation}
  \ddot{\phaseRel} - \nabla^2\phaseRel + \frac{4\nuZero\rhoBG\left(\gS-\gC\right)\left(1-2\nuAmp^2\right)}{\hbar^2}\left(\sin\phaseRel + \frac{\lPar^2}{2}\sin^2(\phaseRel)\right) = 0 \, ,
\end{equation}
which is the equation of motion for a field with potential
\begin{equation}
  V(\phaseRel) = \frac{4\nuZero\rhoBG\left(\gS-\gC\right)\left(1-2\nuAmp^2\right)}{\hbar^2}\left(-\cos\phaseRel + \frac{\lPar^2}{2}\sin^2\phaseRel\right) \, .
\end{equation}
In the above we have defined a mean particle density
\begin{equation}\label{eqn:rhoBG}
\rhoBG = \frac{\bg{\densTot}}{2},
\end{equation}
and the parameter
\begin{equation}
  \lPar^2 = \frac{2\delta^2(\gS-\gC)\rhoBG}{\nuZero} \, ,
\end{equation}
which controls the shape of the effective potential, and can be experimentally tuned.
For $\lPar = 0$, this reduces to the sine-Gordon model, while for $\lPar > 1$, the potential for the effective relativistic scalar develops a series of local minima, as illustrated in~\figref{fig:effective-pot}.
Meanwhile, the Lagrangian has a global $U(1)$ symmetry associated with an overall phase rotation of both condensates, so the total phase is a Goldstone mode and does not develop a potential.
We can therefore identify the relative phase phonons as the appropriate variable in which to study metastability, even in the nonlinear regime. 

\section{Floquet Instabilities and Nonlinear Dynamics of Coupled Condensates}\label{sec:floquet-modes}
Above we briefly outlined the connection between dilute gas cold atom BECs, described by coupled Gross-Pitaevskii equations, and relativistic scalar fields.
For simplicity, we only considered the symmetric case, resulting in equal densities for each condensate in the homogeneous false vacuum state.\footnote{Only $g_{11}=g_{22}$ is required to obtain equal condensate densities, not the additional requirement $\gC=0$ that we have included in our definition of a symmetric theory.}
The equal background densities, in turn, ensured that the total and relative phonons decouple from each other in the linear regime.  
Ignoring backreaction and rescattering of the inhomogeneities, the background dynamics is independent of the fluctuations and acts as an external input to the fluctuation equations.
Therefore, the choice $g_{11}=g_{22}$ ultimately allows us to treat the linear dynamics of the total and relative phonons independently.
As listed in the previous section, a number of assumptions about the condensate dynamics must hold in order for the false vacuum derivation to be valid.
In this section, we will use nonlinear simulations to test these assumptions.
Specifically, we will assume the condensates evolve according to the coupled GPEs, and that initially they are well-approximated as spatially homogeneous.
Given this, the emergence of an effective relativistic scalar field theory posessing an analog false vacuum rests on two key assumptions about the resulting fluctuation dynamics:
\begin{enumerate}[A.]
  \item perturbations in the local number densities are small and can be integrated out at quadratic order in the action, and
  \item all relevant dynamical condensate modes only feel the time-averaged effects of the $\nuT$ modulations.
\end{enumerate}
Further, reducing to a single effective scalar requires a decoupling between the total and relative phase phonons.
The validity of these assumptions for the linearized fluctuation dynamics was studied in detail by Braden \etal~\cite{Braden:2017add}.
We showed that for $\nuAmp \ll 1$, the introduction of periodic modulation in $\nuT$ of frequency $\omega$ induces a band of exponentially unstable modes with wavenumber centered at
\begin{equation}\label{eqn:k-unstable}
  \frac{\hbar^2k^2}{\rhoBG m(\gS-\gC)} \approx 2\left(\sqrt{1+\frac{\hbar^2\nuFreq^2}{4\rhoBG^2(\gS-\gC)^2}}-1\right) - 4\cphi\frac{\nuZero}{\rhoBG(\gS-\gC)} \, ,
\end{equation}
where $\cphi = \cos{\bg{\phaseRel}} = \pm 1$.
This exponential growth occurs in both the relative phase and relative density fluctuations, which if left unchecked will lead to a breakdown of the first two assumptions above.
The emergence of these exponentially growing modes required only that the condensates: 
\begin{enumerate}
  \item followed the coupled GPEs~\eqref{eqn:gpe-cartesian}, 
  \item were subjected to a periodic modulation in $\nuT$, 
  \item could be treated as nearly spatially homogeneous, and
  \item small initial inhomogeneous fluctuations were present.
\end{enumerate}
Subject to these assumptions, the presence of the Floquet band is unavoidable.
The third condition is not strictly necessary to excite Floquet instabilities, but consideration of an inhomogeneous background will significantly modify the scalar field interpretation, so we will not pursue it here.
The fourth condition simply provides the seed for exponential linear growth, and can be fulfilled either by requiring that the initial fluctuations satisfy the uncertainty principle, or by populating the Floquet band through nonlinear interactions described by the coupled GPEs.

The first obstacle to realizing an analog false vacuum is to ensure that either the Floquet instabilities are under control or that their presence does not qualitatively alter the dynamics of the longer wavelength modes essential for bubble nucleation.
One possibility is that the full nonlinear dynamics sequesters the effects of linear instability from the IR modes required to form bubbles.
Below we demonstrate that, by itself, nonlinear evolution with the coupled GPEs does not provide a mechanism to isolate the effects of the Floquet instability from the IR modes.

Therefore, the existence of the analog false vacuum depends on the existence of a physical mechanism in the UV to prevent the Floquet modes from appearing.
Some examples of these corrections include:
\begin{enumerate}
\item corrections to the $S$-wave scattering approximation from atoms retaining memory of the internal structure of the interaction potential between collisions;
\item corrections to the short-wavelength dynamics of the condensate from the finite mode occupancy;
\item the discrete nature of the atoms leading to a breakdown of the continuum field description;
\item interactions of the collective phonon modes with the uncondensed thermal modes of the gas; and
\item corrections to the one-dimensional approximation from excitations of transverse degrees of freedom associated with the trapping potential in the transverse directions.
\end{enumerate}
If the fourth effect is large, then we expect the decay dynamics to be driven by thermal fluctuations, rather than vacuum fluctuations.  Meanwhile, the final effect will lead to strong modifications to the effective relativistic field theory description.  Therefore, the first three effects are the most promising from the viewpoint of analog false vacuum decay.
Although all three of these effects will most strongly modify the dynamics at large wavenumbers, they may also change the IR dynamics of the bubbles.
Therefore, detailed analysis is required to understand any modifications to the false vacuum decay interpretation that results from these corrections to the truncated Wigner description.

There are two potentially relevant UV scales: the healing length of the condensate (the length scale over which the condensate will respond to a defect or boundary) and the interatomic separation. For symmetric condensates, the wave numbers associated with the healing length and interatomic separation are 
\begin{equation}\label{eqn:UV-scales}
k_{\rm heal} = \frac{\sqrt{\gS\rhoBG m}}{\hbar} , \qquad k_{\rm atom} = \frac{8 \pi^2 \hbar^2}{\gS m} \, .
\end{equation}
The validity of the dilute gas BEC approximation requires $k_{\rm heal} \ll k_{\rm atom}$, making the healing length the first relevant UV scale.

Determining if any of these mechanisms is active and sufficient to remove the Floquet modes requires detailed physical modelling of any proposed experimental setup, which we leave to future work.
However, the discrete nature of the atoms provides a promising mechanism to remove the Floquet modes.
The continuum description must break completely at $k_{\rm atom}$, and hence modes above this scale cannot exist within the condensate.
As well, we expect that the finite mean free path of the atoms will induce additional viscous corrections at wavenumbers somewhat below $k_{\rm atom}$, which will preferentially damp large wavenumbers.
Since $k_{\rm heal}$ encodes the scale above which the dispersion relationship changes to that of free particles, these additional viscous corrections may enter at scales not too far removed from the healing length.
Another example of a relevant effect in another experimental setup can be found in \myref~\cite{2018arXiv180707564N}, where modifications of the trap geometry are used to induce ``synthetic dissipation'' of high-k modes and the ability to tune the dissipation scale.
Referring to~\eqref{eqn:k-unstable}, we also see that a degree of experimental tunability exists, since either increasing the modulation frequency $\nuFreq$ or decreasing the condensate density $\rhoBG$ will move the instability to higher wavenumbers relative to the healing length.  However, one must also beware that increasing the driving frequency could cause additional modes (such as transverse excitations in the trap or additional internal excited states of the atoms) to become excited.  If this happens, then the dynamics will differ dramatically from the coupled GPEs discussed here.

\subsection{Nonlinear dynamics}
\begin{figure}
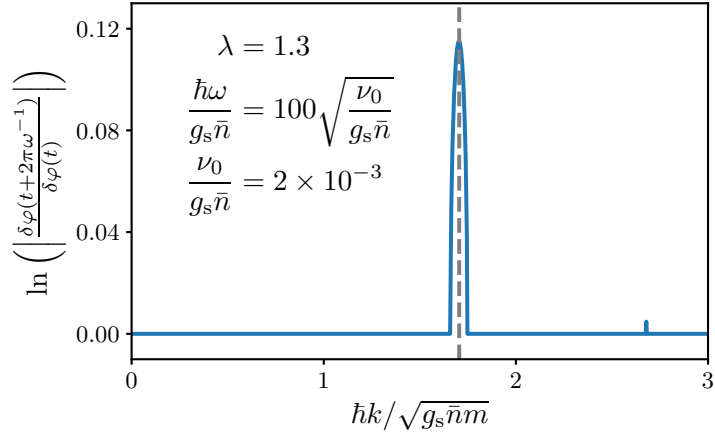
\centering
  \includegraphics[width=\singleWid]{{{floquet-chart-fiducial}}}
  \caption{Lyapunov exponents (\ie the maximal real component of the Floquet exponents) for linear fluctuations in the analog cold atom false vacuum for our fiducial model with ${\lPar=1.3}, {\frac{\nuZero}{\gS\rhoBG} = 2\times 10^{-3}},{\frac{\hbar\nuFreq}{\gS\rhoBG}=100\sqrt{\frac{\nuZero}{\gS\rhoBG}}}$.   The analytic estimate for the wavenumber of the center of the band~\eqref{eqn:k-unstable} is shown as a vertical dashed gray line.  We also see a higher-order band of weaker amplitude that appears at ${\frac{\hbar k}{\sqrt{\gS\rhoBG m}}\sim 2.7}$, which can be obtained from energy-momentum conservation in higher-order tree level diagrams.}
  \label{fig:floquet-band}
\end{figure}

We now study the full nonlinear dynamics of the coupled condensates, properly accounting for the presence of the Floquet band.
For concreteness, throughout this section we choose parameters
\begin{equation}
  \lPar = \delta\sqrt{\frac{2\gS\rhoBG}{\nuZero}} = 1.3, \qquad \frac{\nuZero}{\gS \rhoBG} = 2\times 10^{-3}, \qquad \frac{\hbar\nuFreq}{\gS\rhoBG} = 100\sqrt{\frac{\nuZero}{\gS\rhoBG}} \, ,
\end{equation}
which were suggested as reasonable values by Fialko \etal~\cite{Fialko:2014xba}.
The initial fluctuation amplitudes for the nonlinear simulations are set by the dimensionless particle density
\begin{equation}
  \frac{\hbar\rhoBG}{2\sqrt{m\nuZero}} = 10^{3} \, ,
\end{equation}
again as suggested in \myref~\cite{Fialko:2014xba}.
The maximal Lyapunov exponent for fluctuations around the fiducial false vacuum as a function of dimensionless wavenumber $\kdim = \frac{\hbar k}{\sqrt{\gS\rhoBG m}}$ are shown in~\figref{fig:floquet-band}.
We see a narrow band of unstable wavenumbers centered around $\frac{\hbar k}{\sqrt{\gS \rhoBG m}} \approx 1.705$, as predicted by~\eqref{eqn:k-unstable}.
Note that for these choices of parameters, this is slightly above the UV scale associated with the healing length, so it is not unrealistic to expect corrections to appear.
However, since we wish to understand the implications of assuming the coupled GPEs hold, we proceed without considering any such corrections.
\Figref{fig:floquet-band} provides a sharp prediction for the wavenumbers at which the GPE description must break down in order to maintain the analogy with false vacuum decay.

\begin{figure}[h]
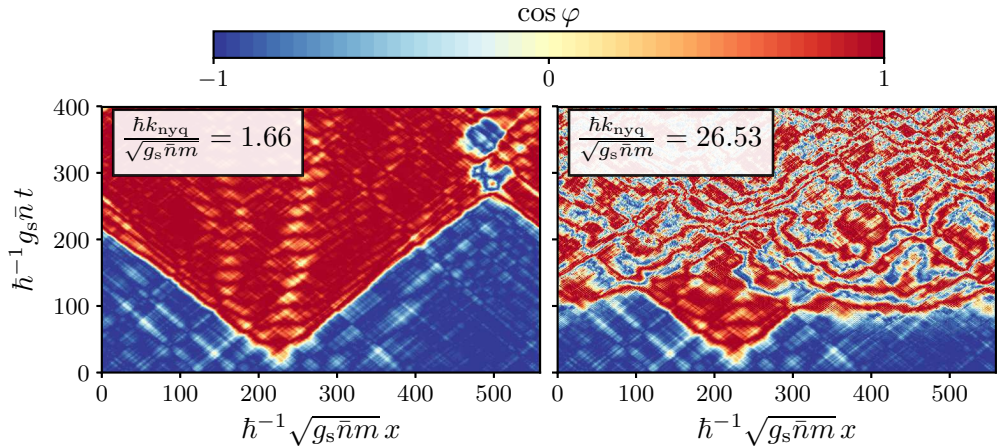

  \includegraphics[width=\fullWid]{{{bubbles-spectral-show-floquet}}}
  \caption{A demonstration of the drastic change in dynamics induced by the eventual nonlinear interactions of the exponentially growing Floquet modes relative to expectations from a time-averaged analysis.  Here we plot the evolution of the cosine of the relative phase $\cos\phaseRel$ for two choices of the lattice spacing (\ie Nyquist frequency) with the initial field profiles and model parameters held fixed.  We have isolated the effects of the Floquet modes by choosing the lattice spacing to either exclude (\emph{left}) or include (\emph{right}) the exponentially unstable modes within the resolved modes of the lattice.   In the left panel ${\frac{\hbar k_{\rm nyq}}{\sqrt{\gS\rhoBG m}} = 1.66}$ is just below the lower edge of the instability band, while in the right panel ${\frac{\hbar k_{\rm nyq}}{\sqrt{\gS\rhoBG m}} = 26.53}$, which is sufficiently far above the upper edge to obtain good convergence of the solution.  Laplacian derivatives were estimated using a Fourier pseudospectral scheme so that the effective lattice wavenumber and continuum wavenumber agree for all resolved modes on the numerical grid.}
  \label{fig:floquet-nonlinear}
\end{figure}

Since the Floquet modes represent a piece of physics beyond that of the time-averaging analysis used to obtain the false vacuum, we expect that the full dynamics will undergo a drastic change as numerical simulation parameters are tuned to either include or exclude their effects.
In~\figref{fig:floquet-nonlinear} we illustrate the full nonlinear dynamics of the relative condensate phase using the numerical procedure described in~\appref{app:trunc-wigner}.
To demonstrate the crucial role the Floquet band has on the dynamics, we have utilized a series of numerical simulations with varying lattice spacings $dx$ (\ie Nyquist wavenumbers) to isolate the effects of the exponentially unstable modes.
The two panels of \figref{fig:floquet-nonlinear} show the evolution of the relative phase for one of these simulations.
The left panel has $\frac{\sqrt{\gS nm}}{\hbar}dx \approx 1.89$ so that $\frac{\hbar k_{\rm nyq}}{\sqrt{\gS n m}} \approx 1.66$, and the lattice cutoff is just below the start of the instability band illustrated in~\figref{fig:floquet-band}.
As a result, the unstable Floquet modes cannot be resolved by our simulation, and we expect the time-averaging analysis to be valid.
Indeed, we see the dynamical nucleation and subsequent expansion of domain wall-antiwall pairs (\ie bubbles in one dimension) in the relative phase, indicating a first-order phase transition.
By contrast, the right panel takes ${\frac{\sqrt{\gS \rhoBG m}}{\hbar}dx \approx 0.12}$ and ${\frac{\hbar k_{\rm nyq}}{\sqrt{\gS\rhoBG m}} \approx 26.53}$, which is above the upper edge of the instability band.\footnote{To ensure that the dynamics of the short-wavelength modes is properly resolved, we have chosen the lattice cutoff small enough to saturate the convergence in both $dt$ and $dx$ (see~\appref{app:convergence}), and ensure the Noether charges are preserved to the $\mathcal{O}(10^{-15})$ level.  To demonstrate the change in behavior occurs from effects localized in the Floquet band, we also ran a simulation with $\frac{\sqrt{\gS nm}dx}{\hbar} \approx 1.79$ and $\frac{\hbar k_{\rm nyq}}{\sqrt{\gS n m}} \approx 1.75$.  This grid just resolves the full Floquet band, and is only a  $\mathcal{O}(10)$\% change in the grid spacing compared to the left panel of \figref{fig:floquet-nonlinear}.  The qualitative behavior matches that of the right panel of \figref{fig:floquet-nonlinear}, with the false vacuum state being lost due the Floquet modes.}
As expected, we see a drastic change in the resulting dynamics of the relative phase.
During the initial stages (while the Floquet modes still have a small amplitude), the evolution matches that of the left panel.
However, at $\frac{\gS\rhoBG}{\hbar} t \sim 100$ the exponentially growing linear modes become $\mathcal{O}(1)$ and undergo strong mode-mode coupling.
The evolution is subsequently dominated by short-wavelength modes of large amplitude, competely erasing the first-order phase transition dynamics.
The Floquet modes therefore qualitatively change the condensate dynamics from that of an effective relativistic scalar field trapped in a false vacuum.
In each simulation, we used the same realization of the initial fluctuations.  In particular, since the lattice cutoff is below the Floquet band in the left panel, this means we did not initially populate the exponentially unstable modes.  Although physically unrealistic since it violates the uncertainty principle, this was done to demonstrate that the nonlinear dynamics of the coupled GPEs will populate these modes.  Since fluctuation power is being dynamically generated rather than input as an initial condition, it also demonstrates the need to have a physical model for the behavior of these modes.  Here we have taken this model to be the coupled GPEs, and shown that this assumption is inconsistent with the existence of a false vacuum decay.

\begin{figure}
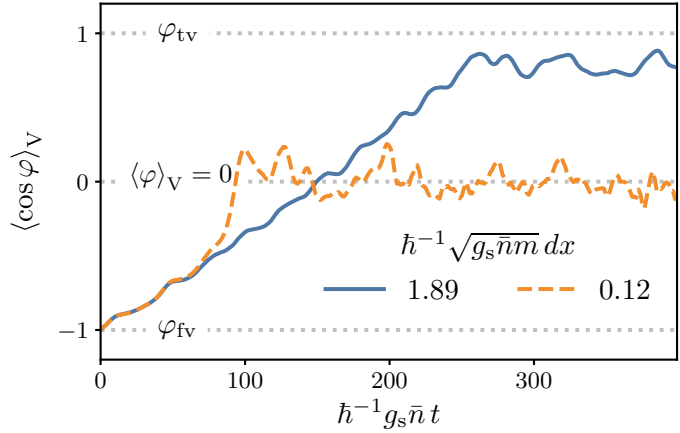
\centering
  \includegraphics[width=\singleWid]{{{mean-cos-phi}}}
  \caption{Evolution of the spatial average $\langle\cos\phaseRel\rangle_{\rm V}$ for the simulations shown in~\figref{fig:floquet-nonlinear}.  The simulation without the unstable Floquet modes is shown in solid blue, and with the unstable modes in dashed orange.  For the false vacuum interpretation, this quantity should evolve from being nearly $-1$ to being close to $1$.  Because of the nonlinear nature of $\cos\phaseRel$, there will be some offset from the true and false vacuum values in the expectation value.  However, while the case without Floquet modes shows clear evidence of the phase transition occuring, this does not occur for the simulations that resolve the Floquet band, where the mean instead goes to roughly $0$.  For reference, the false vacuum ($\phaseRel_{\rm fv}$), true vacuum ($\phaseRel_{\rm tv}$), and random phase ($\langle\phaseRel\rangle_{\rm V}$) results are shown as dotted gray lines.}
  \label{fig:mean-cos-phi}
\end{figure}

This is further demonstrated in~\figref{fig:mean-cos-phi}, where we show the lattice average of $\cos\phaseRel$ for the simulations in~\figref{fig:floquet-nonlinear}.
At the false vacuum $\cos\phaseRel_{\rm fv} = -1$, while in the true vacuum $\cos\phaseRel_{\rm tv}=1$,
so that if the first-order phase transition interpretation holds, the expectation value should go from approximately $-1$ to $1$.
Corrections to the pure $\pm 1$ behavior seen in~\figref{fig:floquet-nonlinear} arise from domain walls and fluctuations present in the condensates.
However, we clearly see that in the absence of Floquet modes this intuition holds, while when the Floquet modes are present the transition to $\langle\cos\phaseRel\rangle_{\rm V} \approx 1$ is lost and instead we end up with $\langle\cos\phaseRel\rangle_{\rm V} \approx 0$.
This latter behaviour is what we expect if the relative phase is randomly distributed.

Having examined the evolution of the relative condensate phase, we now turn to the local density fluctuations.
This will allow us to study two interesting points regarding the validity of the relativistic scalar field interpretation of the condensate dynamics.
First, as outlined in~\secref{sec:effective-scalar}, the scalar field description rests on integrating out the density perturbations to quadratic order to generate a kinetic term for the relative phase phonons.
This step will be invalid if the density fluctuations become large as a result of the Floquet instability.
In the linear regime, relative density perturbations grow commensurate with the relative phase perturbations for the exponentially unstable modes, and we therefore expect that this assumption will be badly violated in the case where Floquet instabilities remain.
Second, in the linear regime the total and relative phase perturbations decouple (when expanded around either the false or true vacuum).
This decoupling property was essential in truncating the total phase dynamics to obtain an effective action for the relative phase alone.\footnote{As mentioned above, the emergence of the relative phase as the key tunneling variable comes directly from the Hamiltonian term proportional to $\nuT$, and thus holds to nonlinear order as well.
Therefore, the decoupling of the relative and total phase phonons allows for a simpler treatment of single field tunneling, as opposed to a more complex two field problem.}
As we show below, in the absence of the Floquet modes, both the small amplitude and decoupling approximations hold throughout the simulation.  However, they are both broken as soon as the dynamics of the Floquet band is considered.

\begin{figure}
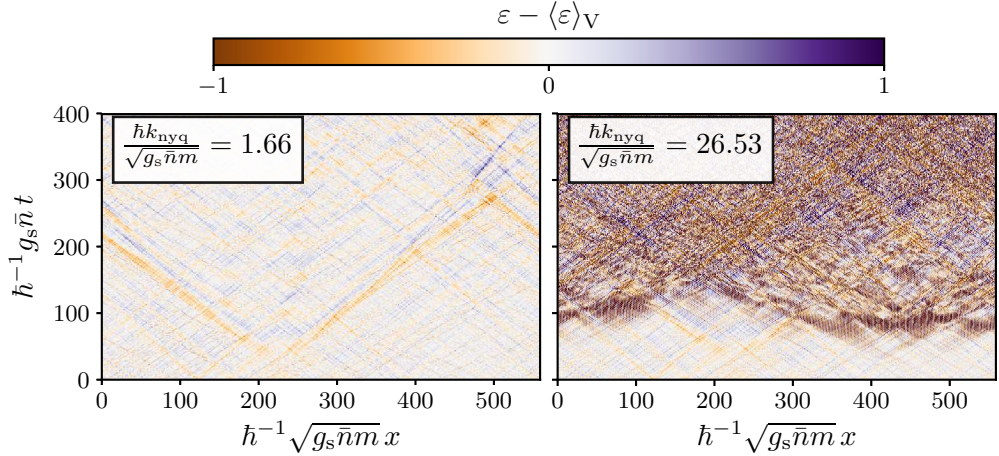

  \includegraphics[width=\fullWid]{{{rho-diff-evolution-floquet}}}
  \caption{Time-evolution of relative density perturbations for the same simulations as~\figref{fig:floquet-nonlinear}.  In the absence of the rapidly growing short-wavelength modes (\emph{left}), we see that the relative density perturbations remain small throughout the evolution.  Additionally, the highly dynamical regions where the relative phase deviates strongly from either the false or true vacuum (i.e. in the domain walls) are visible as coherent spatially localized fluctuations in the relative number density.  Meanwhile, with the rapidly growing short-wavelength modes (\emph{right}), the fluctuations instead become $\mathcal{O}(1)$ and no trace of any bubble-like structures remain.}
  \label{fig:dens-diff-floquet}
\end{figure}

\Figref{fig:dens-diff-floquet} shows the evolution of the relative density between the two condensates for a simulation without (\emph{left}) and a simulation with (\emph{right}) the unstable Floquet band present.
The simulations used correspond to the left and right panels of~\figref{fig:floquet-nonlinear}.
As expected, in the absence of the Floquet instability, the fluctuations remain small throughout the evolution.
We also note that the locations of the bubble walls and collision regions evident in the relative phase also manifest as coherent fluctuations in the relative density.
Meanwhile, when the Floquet modes are present the evolution is dramatically different.
At early times, the evolution matches that of the simulation without the Floquet band.
However, at $\frac{\gS\rhoBG}{\hbar} t \sim 100$, $\mathcal{O}(1)$ fluctuations of short wavelengths dominate the evolution, indicating a breakdown of the interpretation in terms of a relativistic scalar field.

\begin{figure}
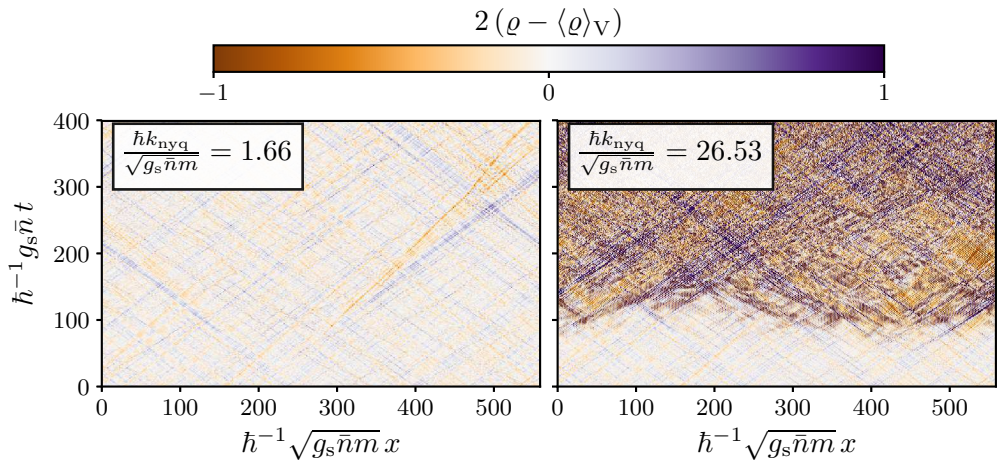

  \includegraphics[width=\fullWid]{{{rho-tot-evolution-floquet}}}
  \caption{Manifestation of the Floquet instability in the total number density for the same pair of simulations as~\figref{fig:floquet-nonlinear}. As before, we see that the presence or absence of the Floquet modes radically alters the behavior of the system.  However, unlike the relative phase, there is no trace of the bubble walls and collision regions, providing an indication of the decoupling of the relative and total density phonons.}
  \label{fig:dens-total-floquet}
\end{figure}

In~\figref{fig:dens-total-floquet} we instead plot the evolution of the total condensate densities.
As with the relative density fluctuations, we see stable behavior in the absence of the Floquet modes.
Unlike the case of the relative density fluctuations, we see no evidence for the bubbles in the evolution of the total density.
This indicates that, in the absence of the exponentially growing modes, the relative and total phonons remain decoupled.
However, as soon as the Floquet dynamics are correctly captured by the simulations, large fluctuations in the total particle density develop as well.
Since the total density phonons are stable in linear perturbation theory, this indicates that the growth of $\densTot$ fluctuations is driven by nonlinear scattering with the growing Floquet modes of $\densRel$.
Thus, the decoupling between the total and relative phonons is destroyed by the presence of the Floquet resonance.

\begin{figure}
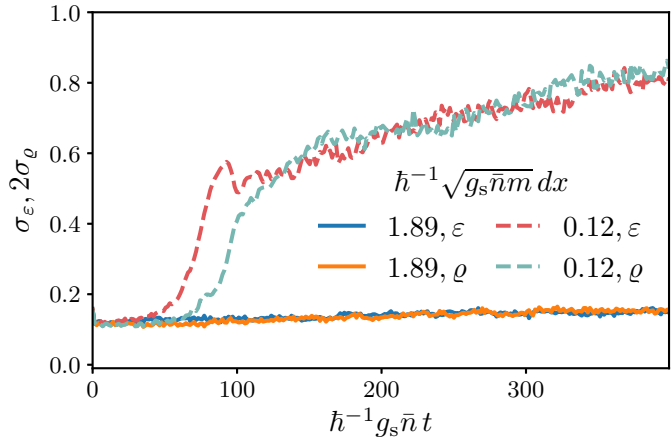
\centering
  \includegraphics[width=\singleWid]{{{std-rho-both}}}
  \caption{Time evolution of the RMS fluctations ($\sigma_{\densRel/\densTot}$) in both the relative (\emph{blue, red}) and total (\emph{orange, green}) number densities of the condensates.  We consider the same simulations as~\figref{fig:floquet-nonlinear}, with the coarse simulation missing the Floquet band given by solid lines, and the finer simulation including the Floquet band shown with dashed lines.  In the absence of the Floquet band, the fluctuations maintain a stable amplitude, while they grow rapidly in the presence of the Floquet instability.  Additionally, the relative density perturbations grow first, directly from the linear Floquet instability.  Subsequently, the total density perturbations begin to grow (at a faster rate) as nonlinear interactions transfer perturbations in the relative phonon sector into the total phonon sector.  The growth rate quenches slightly at $\frac{\gS\rhoBG}{\hbar} t \sim 100$, when the fluctuations begin to interact through strong mode-mode coupling.}
  \label{fig:density-rms}
\end{figure}

Finally, \figref{fig:density-rms} shows the RMS fluctuations in the relative and total density fluctuations, normalized to the (constant) mean total density over the lattice.
This provides an alternative visualization of the breakdown of the effective scalar field description, and subsequent loss of decoupling between the relative and total phonons.
In particular, in the absence of the Floquet instability, the fluctuation amplitude remains stable throughout the course of the simulations.
However, as soon as the Floquet band is present, we see a rapid initial growth in the fluctuations of both the relative and total energy densities.
The growth slows somewhat at $\frac{\gS\rhoBG}{\hbar} t \sim 100$, consistent with partial quenching of the linear instability by nonlinear mode-mode coupling as seen in the real space evolution above.
Further, the growth of fluctuations in the total density slightly lags the growth in the relative density.
Such a lag is consistent with total density fluctuations being induced by nonlinear scattering with the exponentially growing relative density fluctuations.
This further shows that the loss of decoupling between the total and relative phonons occurs because of nonlinear interactions arising from the rapid growth of Floquet modes in the linear regime.

In this section, we have investigated the effects of Floquet instabilities arising in the analog false vacuum decay proposal of Fialko \etal~\cite{Fialko:2014xba,Fialko:2016ggg}, which were originally pointed out in Braden \etal~\cite{Braden:2017add}.
The existence of these instabilities leads to a complete breakdown of the analogy between the BEC evolution and a relativistic scalar field, and thus spoils the false vacuum analogy.
Therefore, the assumption that the condensates obey the coupled GPEs at all scales is inconsistent with the creation of a false vacuum by external periodic modulation of a coupling constant.
Meanwhile, in cases where the unstable modes are artificially removed from the dynamics, the false vacuum interpretation is restored, at least at a qualitative level.
Although this presents an obstacle to successfully realizing an analog false vacuum decay experiment, it is important to remember that the coupled GPEs are an approximation to the full dynamics of the BECs.
In an actual experiment, additional dynamics will be present to modify the GPE description.
From the viewpoint of the successfully realizing false vacuum decay in the current proposal, we require new short-wavelength physics not captured by the GPE to enter at length scales longer than $\lPar_{\rm unstable} = 2\pi k_{\rm unstable}^{-1}$.
This new physics must remove the exponentially growing Floquet modes, while simultaneously exerting a minimal influence on the dynamics of the bubble-forming modes.
Two examples of how this may occur are: damping of the growing modes, and an effective projection of the modes to remove them from long-wavelength condensate dynamics.
A number of physical mechanisms can lead to these types of effective modifications to the coupled GPEs.
Due to the finite interatomic spacing, the continuum description assumed by the GPEs fails completely for wavelengths of order this spacing, guaranteeing all shorter wavelength modes will effectively be projected from the system.
Further, we expect the effects of this spacing and the finite mean free path of the atoms to induce viscous like corrections at somewhat longer wavelengths, similar to the case of a normal fluid, leading to the damping of high-wavenumber modes.
Similarly, interactions between the condensate and the cloud of uncondensed particles will induce damping of condensate modes, which may help to counteract the Floquet instability.
Finally, a particularly intriguing possibility is the induce synthetic dissipation into the system, allowing for a tunable cutoff in the GPE description~\cite{2018arXiv180707564N}.
However, we expect the effects of realistic physical corrections will also leak into the dynamics of the longer wavelength modes needed to nucleate the bubbles.
Therefore, detailed analysis of the nonlinear dynamics is required to understand how the corrections needed to tame the Floquet instabilities modify the interpretation of the bubble nucleation dynamics. We leave such an exploration to future work.

\section{First and Second Order Phase Transitions of the Analog False Vacuum}\label{sec:nonlinear-scalar}
In the previous section we confirmed the presence of linear Floquet instabilities within the GPE description of the analog false vacuum cold atom BEC proposal.
Using nonlinear simulations, we further demonstrated the need to consider physical corrections beyond the use of coupled GPEs and the truncated Wigner approximation in order to preserve the false vacuum interpretation.
We briefly indicated some physical mechanisms that may act to effectively remove the instabilities from the system.
We left the complex issues of detailed modeling of these mechanisms, and how they impact the interpretation in terms of a relativistic scalar to future work.
To help motivate these future studies, we now investigate the nonlinear dynamics under the assumption that we can find an experimental realization where the Floquet modes have been removed and the truncated Wigner approximation is valid, in which case the effective scalar field description should hold.
A range of dynamical symmetry breaking phenomena can be arranged, including both second- and first-order phase transitions.
In each case, the transition is driven by the dynamics of quantum fluctuations, opening the possibility of experimentally testing many fundamental aspects of quantum field theory.

For simplicity, throughout this section we remove the Floquet modes by setting a hard lattice cutoff so that the Nyquist frequency is below the Floquet band.
However, one should be aware that the introduction of the finite lattice cutoff can lead to both a numerical pileup of fluctuation power at the grid scale, and aliasing of short wavelength power into longer wavelengths.
Each of these can significantly modify the nucleation rate of bubbles, and in the former case can lead to a spurious nucleation of bubbles.
This is especially a problem when the cutoff wavenumber is below that required to properly capture the full profile of a nucleated bubble.
To overcome this limitation, in this section we also take the frequency of the external driver $\frac{\hbar\nuFreq}{\gS\rhoBG} = 3200\sqrt{\frac{\nuZero}{\gS\rhoBG}} \approx 143.11$ to be sufficiently large that the Floquet band occurs at higher wavenumbers than those needed for a convergence of the IR modes.
The grid spacing for all simulations is $\frac{\sqrt{\gS\rhoBG m}}{\hbar} = \frac{25}{2048}\sqrt{\frac{\gS\rhoBG}{\nuZero}}$, which is sufficiently converged that the simulation results are indistinguishable by eye as the grid spacing is varied.
This is simply a numerical trick.
In a real experiment one must ensure that choosing a large driving frequency does not excite additional degrees of freedom, such as internal states of the trapped atoms or higher harmonics of the trap.
However, our goal is to highlight some of the interesting behavior that can be obtained in a pristine theoretical environment,
rather than deal with the many subtleties and additional effects that must be accounted for in an actual experiment.

As shown in \myrefs~\cite{Fialko:2014xba,Fialko:2016ggg,Braden:2017add} and outlined in~\secref{sec:effective-scalar}, the temporal modulation of $\nuT$ creates an effective potential for the long-wavelength modes of $\phaseRel$
\begin{equation}
  \label{eqn:rel-phase-pot}
  V_{\rm eff}(\phaseRel) =  V_0\left(-\cos\phaseRel + \frac{\lPar^2}{2}\sin^2\phaseRel\right) \, ,
\end{equation}
illustrated in~\figref{fig:effective-pot}, where the parameter
\begin{equation}
  \lPar = \nuAmp\sqrt{\frac{2\gS \rhoBG}{\nuZero}}
\end{equation}
is experimentally tunable.
For $\lPar \leq 1$, $\phaseRel=\pi$ is a local maximum, while it is a local minimum for $\lPar > 1$.
In the (time-averaged) scalar field interpretation of $\phaseRel$, increasing $\lPar$ adjusts the field from sitting on top of a hill to sitting in a local minimum.
For initial states localized around the false vacuum $\phaseRel = \pi$, we expect two distinct mechanisms by which the false vacuum state can decay.
The value of $\lPar$ determines which mechanism is relevant.

For $\lPar < 1$, modes with ${k \lesssim 2\frac{\sqrt{\nuZero\gS\rhoBG}}{\hbar}\sqrt{1-\lPar^2}}$ will experience a tachyonic instability, 
We confirmed this at the level of linear perturbation theory in \myref~\cite{Braden:2017add}, where we also numerically obtained $\frac{\nuZero}{\gS\rhoBG}$ and $\nuAmp$ dependent corrections.
Based on this picture, we expect that for $\lPar < 1$, $\phaseRel$ falls off the top of the hill through a spinodal instability.
This leads to the rapid creation of a set of domains (with the field localized near either the $\phaseRel = 0$ or $\phaseRel = 2\pi$ vacuum) connected by a network of domain walls.
Meanwhile, for $\lPar > 1$ the local maxima become local minima of the potential, corresponding to false vacuum states.
In this regime, we obtain a first-order phase transition and the false vacuum decays via the nucleation of bubbles.
Of course, we do not necessarily expect a sharp distinction between the second-order and first-order phase transitions to occur at $\lPar \approx 1$.
In particular, when $0< \lPar - 1 \ll 1$, the barrier between the false and true vacuum is narrow and shallow.
Therefore, we expect that in some spatial regions the initial condensate fluctuations will probe over the barrier of the potential and fall towards the true vacuum.
For very shallow barriers, these transitions happen in a significant fraction of the volume, and rather than having a few well-defined bubble nucleations, many regions of the true vacuum appear simultaneously.
Such dynamics is intermediate between the tachyonic growth of long-wavelength modes and the regime of rare well-defined bubble nucleations.

\Figref{fig:stabilise-vary-delta} illustrates the nonlinear stabilization of the low-$k$ tachyonic modes by increasing the modulation amplitude $\nuAmp$ of the interspecies conversion rate.
The top two panels have $\lPar < 1$.
We observe the rapid emergence of a dynamically evolving domain wall network, as expected from the tachyonic growth of the long-wavelength modes.
The transitionary behaviour between first- and second-order phase transitions is seen in the middle panels of~\figref{fig:stabilise-vary-delta}.
Finally, well-defined bubble nucleations appear for $\lPar$ sufficiently larger than one, as seen in the bottom panels of the figure.

\Figref{fig:stabilise-vary-delta} also illustrates the dependence of the false vacuum decay rate on the experimentally tunable parameter $\lPar$.
This effect appears in the figure as a delay in the time at which bubbles nucleate as the value of $\lPar$ is increased,\footnote{This is clear in~\figref{fig:stabilise-vary-delta} since we start from the same realization of the initial fields in each case.  In the more general case, where the initial conditions are also sampled, then this effect would only be evident upon considering an ensemble of time-evolutions, rather than direct comparison of individual simulations.}
and allows for a tuning of typical decay time relative to the expected experimental lifetime of the condensate.
Additionally, the decay rate is sensitive to the number density of condensed atoms, which changes the amplitude of fluctuations in $\wf_i$ relative to the absolute mean.  Since the RMS amplitude of fluctuations is set by $\hbar$ through the uncertainty principle, this is analogous to adjusting the value of $\hbar$. This may provide a window to adjust the expected decay rate while holding the effective scalar potential (\ie the scalar field theory) fixed.

\begin{figure}[h]\centering
  \includegraphics[width=\fullWid]{{{gpe-varyl-multipanel}}}
  \caption{Stabilization of the relative phase of the two condensates as we increase the modulation amplitude $\nuAmp$ and thus $\lPar = \nuAmp\sqrt{2\gS\rhoBG/\nuZero}$.  Here we plot the time evolution of $\cos\phaseRel$ for a series of runs with varying choices of $\lPar$.  In the false vacuum we have $\cos\phaseRel = -1$, while $\cos\phaseRel = 1$ in any of the true vacua.  From top left to bottom right, we have $\lPar = 0,0.9,1,1.1,1.3$, and $1.5$.  We see the transition from a spinodal instability (for $\lPar \ll 1$) into a regime of metastability dominated by bubble nucleation for $\lPar \gg 1$, with a corresponding transitionary regime around $\lPar \sim 1$.}
  \label{fig:stabilise-vary-delta}
\end{figure}

\subsection{Validity of Scalar Field Interpretation and Phonon Decoupling}
As we saw in \secref{sec:floquet-modes}, when Floquet instabilities induced by the external driving of the system are present both the phase and local number density fluctuations grow rapidly.
This invalidates the assumption that density fluctuations are small, which is needed to derive the effective scalar field description.
We now investigate whether or not a similar breakdown occurs when the short-wavelength Floquet modes have been excised from the evolution, but an exponential instability remains in the long-wavelength modes.
In particular, in the regime of spinodal instability ($\lPar < 1$), the exponentially growing long-wavelength modes that destabilize the local maximum are simply another band of exponentially unstable linear modes.
It is therefore important to explicitly check the validity of various approximations in this regime.

\begin{figure}
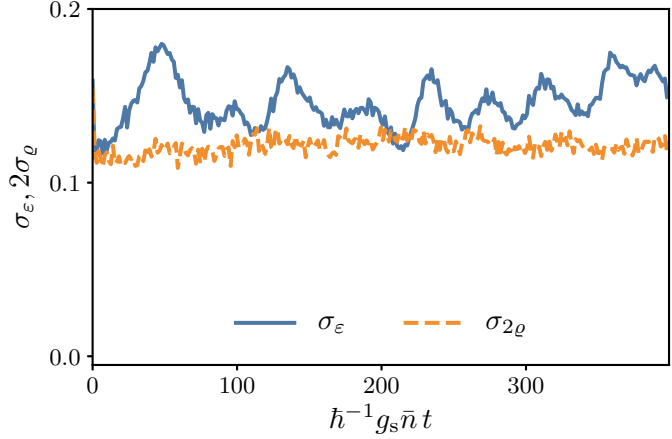
\centering
  \includegraphics[width=\singleWid]{{{density-rms-l0}}}
  \caption{The evolution of the RMS of the relative (\emph{blue solid}) and total (\emph{orange dashed}) density perturbations for the $\lPar=0$ simulation in~\figref{fig:stabilise-vary-delta}.  Although initially the amplitude of the relative density perturbations grow, they saturate before becoming strongly nonlinear.  The total density perturbations, on the other hand, maintain a stable amplitude throughout the simulation.}
  \label{fig:density-rms-l0}
\end{figure}

\Figref{fig:density-rms-l0} shows the evolution of the RMS of both the total ($\densTot$) and relative ($\densRel$) energy densities for the $\lPar = 0$ simulation.
Although initially the relative RMS of the perturbations grows, unlike in the previous section it saturates before becoming order one.
Further, the fluctuations in the total density remain fairly stable throughout the simulation.
It thus appears that the effective relativistic scalar field interpretation of these simulations will continue to hold (at least at leading order), 
and that the total and relative phonons remain decoupled for the entire evolution.

\begin{figure}
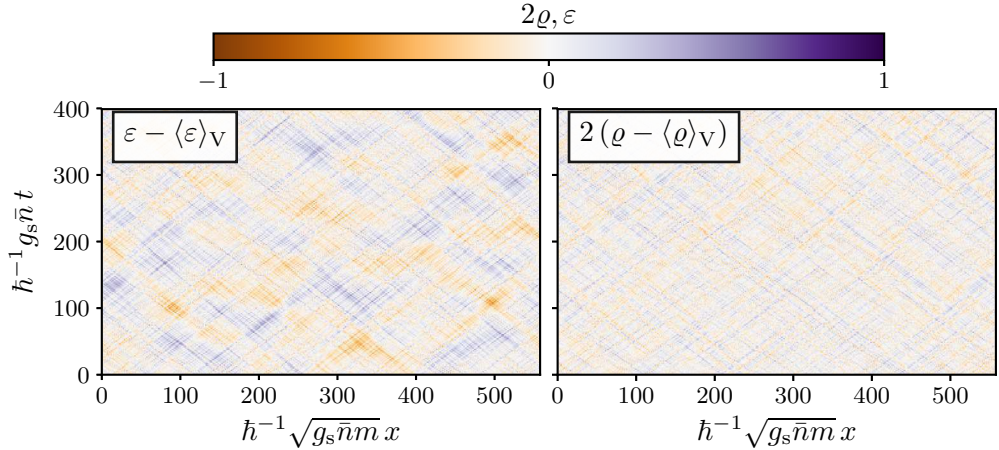

  \includegraphics[width=\fullWid]{{{rho-evolution-l0}}}
  \caption{Evolution of the relative (\emph{left}) and total (\emph{right}) energy densities for the $\lPar = 0$ simulation in~\figref{fig:stabilise-vary-delta}.  We see evidence of the complex domain wall network that develops in the relative phase in the relative density perturbations.  However, there is no apparent signature of this dynamics in the total density fluctuations, indicating that the relative and total phonons remain decoupled in this case.}
  \label{fig:density-evolution-l0}
\end{figure}

This is further illustrated by the evolution of the relative and total densities within the same $\lPar = 0$ simulation, seen in~\figref{fig:density-evolution-l0}.
Comparing to~\figref{fig:floquet-nonlinear}, we see that the density perturbations remain significantly smaller than in the case of excitations in the higher order instability band.
Additionally, large coherent structures in the relative density appear, presumably due to the numerous coherent domain walls in the relative phase.
However, no such structures appear in the total density fluctuations.
Combined with the lack of growth in these fluctuations overall, this suggests that the total phase phonons remain decoupled from the relative phase phonons in this case as well.

\section{Conclusions}\label{sec:conclusions}
In this paper, we investigated the nonlinear stability of analog false vacuum decay experiments in cold atom Bose-Einstein condensates.  
In particular, we first showed that using the coupled Gross-Pitaevskii equations as a physical model for the evolution of the condensates is inconsistent with the generation of an effective false vacuum through modulation of the interspecies conversion rate.
To do this, we confirmed the presence of the linear Floquet instabilities studied in \myref~\cite{Braden:2017add} in full nonlinear simulations.
We then demonstrated that excitation of these modes leads to a complete breakdown of the effective scalar field description, and therefore the false vacuum decay analogy.
Since the exponentially unstable modes appear at nonzero wavenumber, it is possible that their effects can be removed from the system if the Gross-Pitaevskii equations used to model the condensates break down at wavelengths above the instability scale.
Some of these possible mechanisms were briefly outlined in section 4.
However, this requires modeling of the condensates beyond the coupled GPEs, and such corrections must be analyzed on a case-by-case basis to determine whether a given experiment is viable.  Further, the original analogy between the relativistic scalar field and cold atom dynamics was obtained assuming the coupled GPEs were the correct description.  Therefore, the mapping of any corrections to the coupled GPEs into the scalar field dynamics must also be understood in order to make quantitative comparisons.  Given these complexities, we did not pursue a full study here.  Rather, our results should be used as an indication that care must be exercised when developing analog cosmological systems using cold atoms. 

If feasible, an analog false vacuum system would provide a unique opportunity to experimentally study fundamental issues in both cosmology and relativistic quantum field theory. Under the assumption that Floquet modes can be removed, we investigated the variety of behavior accessible to experiments by tuning the amplitude of the oscillating interspecies conversion rate. This parameter controls the shape of the potential for the relative phase, taking a local potential maximum to a local potential minimum. In the absence of temporal modulation, if the field begins at the local potential maximum, we demonstrated that the long-wavelength modes of the condensates experience tachyonic growth, and the initial homogeneous state decays via a spinodal instability. Meanwhile, well above this critical amplitude we observed the stabilization of the long-wavelength tachyonic modes and the decay of the metastable state through the formation of domain wall-antiwall pairs (\ie one-dimensional bubbles). The transition between these two regimes is not sharp, but instead near the critical amplitude the decay occurs in an intermediate regime posessing partial characteristics of both the spinodal and bubble nucleation instabilities. In each of these regimes, we found that removing the short-wavelength Floquet instability was sufficient to ensure the density perturbations remain small and the approximate relativistic scalar field description remains valid. The study of this cross-over behavior is one example of the type of physical phenomena accessible to the cold atom systems studied in this paper. 

Our results show the need for detailed physical modelling of the condensate dynamics in order to find an experimental system to realize analog false vacuum decay. In future work, we will investigate corrections to the GPE in specific experimental scenarios in order to determine if the necessary conditions to remove the Floquet instabilities can in principle be met.  We will also determine the detailed mapping (should one exist) to the analog scalar field theory. Theoretical work is also necessary to narrow down the specific observables within the cold atom system (\eg correlation functions) that will be most useful for shedding light on theoretical aspects of vacuum decay. Vacuum decay is both strongly non-linear and strongly quantum mechanical, making it highly likely that we will learn something truly new from analog false vacuum decay. We believe this is strong motivation for future investigation.

\acknowledgments
JB and HVP were supported by the European Research Council (ERC) under the European Community's Seventh Framework Programme (FP7/2007-2013)/ERC grant agreement number 3006478-CosmicDawn.
JB is supported by the Simons Foundation Origins of the Universe program (Modern Inflationary Cosmology collaboration).
The work of JB, HVP, and AP was partially performed at the Aspen Center for Physics, which is supported by the National Science Foundation grant PHY-1607611.
The participation of JB, HVP, and AP at the Aspen Center for Physics was supported by the Simons Foundation.
MCJ is supported by the National Science and Engineering Research Council through a Discovery grant. Research at Perimeter Institute is supported by the Government of Canada through the Department of Innovation, Science and Economic Development Canada and by the Province of Ontario through the Ministry of Research, Innovation and Science.
AP was supported by the Royal Society.
SW acknowledges financial support provided by the Royal Society University Research Fellow (UF120112),  Royal Society Enhancement Grants (RGF/EA/180286,RGF/EA/181015), EPSRC Project Grant (EP/P00637X/1), and partial support from STFC consolidated grant No. ST/P000703/.
This work was partially enabled by funding from the University College London (UCL) Cosmoparticle Initiative.

\bibliography{nonlinear-bec}
\bibliographystyle{JHEP}

\appendix

\section{Stochastic Semiclassical Simulations: The Truncated Wigner Approximation}\label{app:trunc-wigner}
In this appendix we briefly outline our methodology to solve for the dynamics of a two-component cold atom BEC described by the coupled GPEs.
The method, when applied to the GPE, is widely used in the cold atom community and known as the truncated Wigner approximation~\cite{Gardiner:2002}.
More cosmologically oriented readers will recognize the spiritual similarity with semi-classical scalar field lattice simulations used to study preheating~\cite{Felder:2000hq,Frolov:2008hy,Easther:2010qz,Huang:2011gf,Sainio:2012mw}.
The basic idea is to model quantum fluctuations by sampling realizations of the initial quantum state as encoded in the Wigner functional.
The subsequent complex dynamical evolution is then treated by solving the classical equations of motion.
This amounts to propagating a classical probability distribution functional on (very high-dimensional) field space via the method of characteristics.
Given this, it is natural to  interpret the results of a single simulation as the outcome of a single experimental realization.
This approach is known to be exact when the field dynamics is linear and the initial state is Gaussian.
In more complicated nonlinear situations, it can sometimes capture the crucial elements of the dynamics, although the validity must be checked on a case-by-case basis.

Initially, the condensates $\wf_i$ are assumed to be of the form
\begin{equation}
  \wf_i({\bf x}) = \bar{\wf}_i + \delta\hat{\wf}_i({\bf x}) \, ,
\end{equation}
where the $\bar{\wf}_i$ are background solutions, here taken to be stationary points of the coupled GPEs in the spatially homogeneous limit.\footnote{The total phase can experience a linear time-evolution at the stationary points, but we include the background chemical potential $\mu_{\rm bg}$ to remove this.}
Quantum corrections to the ``classical'' background $\bar{\wf}_i$ are modeled by a realization of a Gaussian random field
\begin{equation}\label{eqn:psi-fluc}
  \delta\hat{\wf}_i = \frac{1}{\sqrt{V}}\sum_k\sqrt{\mathcal{P}_i(k)}\hat{a}_ke^{i{\bf k}\cdot{\bf x}} \, ,
\end{equation}
where the $\hat{a}_k$ are draws of uncorrelated complex Gaussian random deviates with unit variance $\left\langle\abs{\hat{a}_k}^2\right\rangle = 1$, and $V$ is the simulation volume.
We generate our samples $\hat{a}_k$ using the Box-Mueller transform
\begin{equation}
  \hat{a}_k = \sqrt{-\ln(\hat{A}_k)}e^{2\pi i\hat{P}_k} \, ,
\end{equation}
where $\hat{P}_k$ and $\hat{A}_k$ are uniform random deviates on the interval $[0,1]$.
Since the fluctuations are assumed to be Gaussian, the spectra $\mathcal{P}_i$ should match the initial quantum 2-point statistics\footnote{An obvious extension allows this to be extended to incorporate correlations between the condensates.}
\begin{equation}
  \langle\delta\hat{\wf}_i({\bf k})\delta\hat{\wf}_j({\bf k'})\rangle_{\rm Q} = \mathcal{P}_i(k)\delta_{ij}\delta({\bf k}-{\bf k'}) \, .
\end{equation}
In an experiment, this will depend on the preparation of the initial state; while in a purely theoretical study it should be adjusted to accurately reflect the physics under investigation.
In particular, these initial fluctuations must be mapped into those of the the relativistic scalar fields in order to correctly interpret the results in the effective scalar field theory.
Some care is need to choose the correct initial fluctuations to reproduce the initial Minkowski state for the effective scalar field $\phaseRel$.
To make contact with the previous literature, here we follow the approach of Fialko \etal~\cite{Fialko:2014xba,Fialko:2016ggg} and choose $\mathcal{P}(k) = \mathcal{A}\Theta(k_{\rm cut}-\abs{\kVec})$ for constant $\mathcal{A}$, resulting in a filtered white noise spectrum for $\delta\hat{\wf}_k$.
Considering the nonlinear transformation into the density and phase variables, we see that this does not correspond precisely to the standard Minkowski vacuum of the effective relativistic scalar field.  We leave consideration of how this impacts the decay rate, and the appropriate initial state to properly simulate the false vacuum, to future work.

From this sampled initial configuration, the fields $\wf_i$ are evolved using the coupled GPEs.
For our numerical scheme, we use a tenth-order accurate Gauss-Legendre integrator (see e.g.~\myrefs~\cite{Braden:2014cra,Butcher:1964}) for the temporal evolution, and a collocation based Fourier pseudospectral approach~\cite{boyd01} for spatial discretization.
When useful, we also compare with a second-order accurate finite differencing approach for the spatial discretization, with
\begin{equation}
  \nabla^2f(x_i) \to L[f](x_i) \equiv \frac{1}{2dx^2}\left(f(x_{i+1})-2f(x_i)+f(x_{i-1})\right)
\end{equation}
and
\begin{equation}
  (\nabla f(x_i))^2 \to G[f](x_i) \equiv \frac{1}{dx^2}\left[\left(f(x_{i+1})-f(x_i)\right)^2 + \left(f(x_{i-1})-f(x_i)\right)^2\right]
\end{equation}
for an arbitrary scalar function $f$.
In all cases, we enforce periodic boundary conditions, explicitly in the case of a finite-difference stencil, and implicitly when using a Fourier pseudospectral method.

\section{Dimensionless Variables}\label{app:dim-var}
In this appendix we briefly present the dimensionless form of our evolution equations and summarize the most important dynamical timescales.
It is convenient to define dimensionless time, space, and condensate variables (here denoted by a bar),
\begin{equation}
  \nodim{t} = \tScl t, \qquad \nodim{x} = \xScl x, \qquad {\rm and} \qquad \nodim{\wf} = \sqrt{\rhoScl}\wf,
\end{equation}
through the introduction of three numerical scales $\tScl$, $\xScl$ and $\rhoScl$ with dimensions of $T^{-1}$, $L^{-1}$, and $L^{d}$ respectively.
It is also convenient to introduce the numerical sound speed ($\csNum$) and energy ($\enScl$),
\begin{equation}
  \csNum \equiv \frac{\tScl}{\xScl} \qquad {\rm and} \qquad \enScl = \hbar\tScl \, .
\end{equation}
Expressed in these variables, the dimensionless Hamiltonian is
\begin{equation}
  \nodim{H} = \frac{1}{\rhoScl\xScl^d}\int \ud{\nodim{x}}{d}\frac{1}{\hbar\tScl}\left(\frac{(\hbar\xScl)^2}{2m_i}\abs{\nodim{\nabla}\nodim{\wf}_i}^2 + \frac{g_{ij}}{2\rhoScl}\abs{\nodim{\psi}_i}^2\abs{\nodim{\psi}_j}^2 - \frac{\nuT_{ij}}{2}\left[\nodim{\wf}_i\nodim{\wf}_j^* + \nodim{\wf}_i^*\nodim{\wf}_j\right]\right)  \, .
\end{equation}
The (dimensionless) combination $\rhoScl\xScl^d$ sets the overall scale of the numerical Hamiltonian, and thus does not enter directly into the evolution equations.
It does, however, set the overall scale of the initial dimensionless fluctuations $\delta\bar{\wf}_i$.
Meanwhile, the numerical energy $\hbar\tScl$ can be used to transform each of the remaining dimensionful couplings into a dimensionless coupling constant.

Alternatively, working directly with the equations of motion, we have
\begin{equation}
  i\frac{\dd \nodim{\wf}_i}{\dd \nodim{t}} = -\frac{\hbar\xScl^2}{2m_i\tScl}\nodim{\nabla}^2\nodim{\psi}_i + \frac{g_{ij}}{\hbar\tScl\rhoScl}\abs{\nodim{\wf}_j}^2\nodim{\wf}_i - \frac{\nuT_{ij}}{\hbar\tScl}\nodim{\wf}_j = -\frac{\nodim{p}_i^2}{2}\nodim{\nabla}^2\nodim{\wf}_i + \nodim{g}_{ij}\abs{\nodim{\wf}_j}^2\nodim{\wf}_i - \nodim{\nuT}_{ij}\nodim{\wf}_j\, .
\end{equation}
We have introduced the dimensionless numerical coefficients
\begin{equation}
  \nodim{p}_i^{2} \equiv \frac{\hbar^2\xScl^2}{m_i\hbar\tScl} = \frac{\hbar\tScl}{m_i\csNum^2}, \qquad \nodim{g}_{ij} \equiv \frac{g_{ij}}{\hbar\tScl\rhoScl}, \qquad {\rm and} \qquad \nodim{\nu}_{ij} \equiv \frac{\nu_{ij}}{\hbar\tScl} \, .
\end{equation}
For convenience, we now specialize to the two condensate case, and define
\begin{equation}
  \gS \equiv \frac{g_{11}+g_{22}}{2},
  \qquad \gD \equiv g_{22}-g_{11}, \qquad
  \gC = g_{12}, \qquad
  \nuT = \nu_{12}, \qquad {\rm and} \qquad
  m \equiv \frac{m_1+m_2}{2}\, .
\end{equation}
For our numerical simulations, we scale the condensates by their average background densities $\rhoBG$, set the numerical sound speed to be the propagation speed of the relative phase phonons, and the dimensionless nonlinear potential coupling to unity
\begin{equation}
  \frac{\abs{\aveV{\bar{\wf}_1}}^2+\abs{\aveV{\bar{\wf}_2}}^2}{2} = 1,
  \qquad \csNum = c_{\rm sound} = \sqrt{\frac{\gS\rhoBG}{m}}, \qquad {\rm and} \qquad
  \gSScl = 1 \, .
\end{equation}
We immediately see
\begin{equation}\label{eqn:paper-norm}
  \rhoScl = \rhoBG^{-1}, \qquad
  \hbar\tScl = \gS\rhoScl^{-1} = \gS \rhoBG, \qquad {\rm and} \qquad
  \hbar\xScl = \sqrt{\frac{\gS m}{\rhoBG\rhoScl^2}} = \sqrt{\gS\rhoBG m} \, ,
\end{equation}
so that the condensates are measured in units of their background density~\eqref{eqn:rhoBG}, and positions in units of the healing length (see~\eqref{eqn:UV-scales}).
The resulting dimensionless equations of motion suitable for numerical simulations are
\begin{subequations}
\begin{align}
  i\frac{\dd\bar{\wf}_1}{\dd\bar{t}} &= -\frac{1}{2}\left(\frac{m}{m_1}\right)\bar{\nabla}^2\bar{\wf}_1 + \left(1 - \frac{\gDScl}{2}\right)\abs{\bar{\wf}_1}^2\bar{\wf}_1 + \gCScl\abs{\bar{\wf}_2}^2\bar{\wf}_1 - \bar{\nuT}\bar{\wf}_2 \\
  i\frac{\dd\bar{\wf}_2}{\dd\bar{t}} &= -\frac{1}{2}\left(\frac{m}{m_2}\right)\bar{\nabla}^2\bar{\wf}_2 + \left(1 + \frac{\gDScl}{2}\right)\abs{\bar{\wf}_2}^2\bar{\wf}_2 + \gCScl\abs{\bar{\wf}_1}^2\bar{\wf}_2 - \bar{\nuT}\bar{\wf}_1 \, .
\end{align}
\end{subequations} 

The commutator for the scaled field variables is
\begin{equation}
  [\bar{\wf}_i(\bar{x}),\bar{\wf}_j^\dagger(\bar{x}')] = \rhoScl\xScl^{d}\delta_{ij}\delta(\bar{x}-\bar{x}')
\end{equation}
which we can write on a discrete lattice with lattice sites $\nodim{x}_m$
\begin{equation}
  [\nodim{\wf}_i(\nodim{x}_m),\nodim{\wf}_j^\dagger(\nodim{x}_n)] = \frac{\rhoScl\xScl^{d}}{d\nodim{x}^d}\delta_{ij}\delta_{mn} = N_{\rm lat}^d\frac{\rhoScl\xScl^{d}}{\bar{L}^d}\delta_{ij}\delta_{mn} \, .
\end{equation}
For our particular normalization of the fields~\eqref{eqn:paper-norm}, the relative amplitude of dimensionless fluctuations in the condensate scales with the total number of condensed particles as $N^{-1/2}$,
\begin{equation}
  [\nodim{\wf}_i(\nodim{x}_m),\nodim{\wf}_j^\dagger(\nodim{x}_n)] = \frac{N_{\rm lat}^d}{N}\delta_{ij}\delta_{mn} \, .
\end{equation}
Since the commutator sets the overall amplitude of vacuum fluctuations through the uncertainty relation, the relative amplitude of density fluctuations to the density in the zero mode scales as $N^{-1}$.

\subsection{Comparison with Dimensionless Units of Fialko \etal}
To ease comparison with previous work, here we also present the dimensionless units used in Fialko \etal~\cite{Fialko:2014xba,Fialko:2016ggg} and translate into the notation of our paper.
To avoid notational confusion, we denote the various normalization constants of Fialko \etal\ with superscript ${}^{\rm F}$ and the corresponding dimensionless quantites by $\nodim{\cdot}^{\rm F}$.
As above, the dimenionless units in our paper are denoted by an overbar $\nodim{\cdot}$.
In those papers, they continue to set the numerical sound speed to the propagation speed of the relative phase fluctuations.
However, they measure time in units of the oscillation frequency $\omega_0 = 2\frac{\sqrt{\nuZero\gS\rhoBG}}{\hbar}$ of the effective scalar field associated with the relative phase, rather than choosing to measure position in units of the healing length.
The condensate wavefunctions are further normalized to the inverse numerical volume scale $\left(\kappa_{\rm F}^{\rm F}\right)^{-d}$.
The corresponding numerical scales are related as
\begin{equation}\label{eqn:fialko-norm}
  \omega_{\rm P}^{\rm F} = 2\sqrt{\nodim{\nuZero}}\omega_{\rm P}, \qquad
  \kappa_{\rm P}^{\rm F} = 2\sqrt{\nodim{\nuZero}}\kappa_{\rm P}, \qquad {\rm and} \qquad
  \Lambda^{\rm F} = \left(\kappa_{\rm P}^{\rm F}\right)^{-1} = \Lambda\frac{\hbar\rhoBG}{2\sqrt{\nuZero m}} \, .
\end{equation}
Finally, they measure $\nuZero$ in units of $\gS\rhoBG$ instead of $\hbar\omega_{\rm P}^{\rm F}$, so that
\begin{equation}
  \nodim{\nuZero}^{\rm F} = \nodim{\nuZero} \, .
\end{equation}

\section{Demonstration of Linearity of Floquet Modes and the Effective Discrete Lattice Wavenumber}\label{app:effective-k}
\begin{figure}[h]
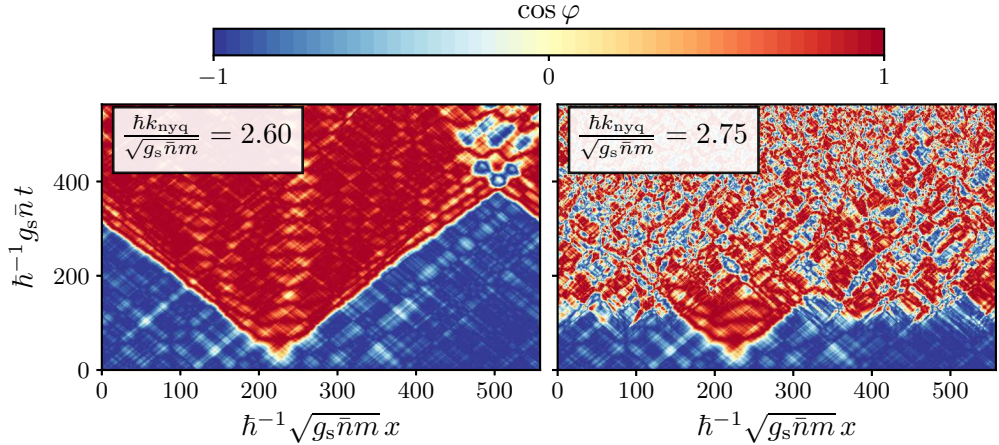

  \includegraphics[width=\fullWid]{{{bubbles-discrete-show-floquet}}}
  \caption{Analogous simulations to the left and right panels of~\figref{fig:floquet-nonlinear}, except using a finite-differencing approximation for the Laplacian instead of a Fourier pseudospectral approximation.  In the left panel we take the Nyquist frequency ${\frac{\hbar k_{\rm nyq}}{\sqrt{\gS\rhoBG m}} = 2.60}$ so that the effective linear lattice frequency ${k_{\rm eff}= \frac{2}{\pi}k_{\rm nyq} \approx 1.66}$ is just below the Floquet instability band.  Meanwhile, in the right panel we take ${\frac{\hbar k_{\rm nyq}}{\sqrt{\gS\rhoBG m}} = 2.75}$, with effective linear lattice frequency ${k_{\rm eff} \approx 1.75}$ just encompassing the full Nyquist band.  Here we do not consider a fully resolved simulation to explicitly illustrate that the breakdown of bubble nucleation description is associated solely with the inclusion of the Floquet band.  Comparing the two panels, we clearly see that the short-wavelength dynamics is associated with the effective frequency of linear fluctuations on the lattice shown in~\figref{fig:effective-k}.}\label{fig:floquet-nonlinear-discrete}
\end{figure}

In~\secref{sec:floquet-modes} we demonstrated the presence of new short-wavelength dynamics that were neglected in previous studies of analog false vacuum decay.
This new dynamics destroyed the effective scalar field description, and also resulted in a loss of the phase transition as measured by the mean value of $\cos\phaseRel$.
Since this dramatic change occurred at precisely the wavenumbers predicted by a Floquet analysis of the linear perturbations, we concluded that this was driven by the Floquet instability.
We now explicitly demonstrate that the new short-wavelength dynamics are indeed a result of a linear instability.
To do this, we rerun the simulations in~\figref{fig:floquet-nonlinear} using a second-order accurate centered finite-differencing approximation for the Laplacian instead of a Fourier pseudeospectral approximation.
The resulting evolution of the relative phase is shown in~\figref{fig:floquet-nonlinear-discrete}.

As outlined in the subsection below, the effective wavenumber felt by linear fluctuations is determined by the choice of discrete Laplacian.
In general it differs from the continuum value as shown in~\figref{fig:effective-k}.
In particular, for the Nyquist mode, the effective linear wavenumber for the second-order accurate Laplacian stencil is $\frac{2}{\pi}$ times the continuum wavenumber.
Since Floquet instabilities arise from linear dynamics, on the lattice the Floquet modes will feel this effective wavenumber.
In contrast, nonlinear interactions not involving derivatives are sensitive to the continuum wavenumber of the fluctuations, and therefore they are not directly influenced by the choice of spatial discretization.
Therefore, if the new dynamics seen with decreasing lattice spacing were associated with nonlinear interactions amongst the fluctuations, it would appear at the same Nyquist frequency, regardless of the choice of Laplacian stencil.
For linear fluctuations, the dynamics will instead appear at different wavenumbers determined by the effective linear wavenumber associated with the choice of discrete Laplacian.
As seen in~\figref{fig:floquet-nonlinear-discrete}, the emergence of the small-scale instability is sensitive to the effective lattice wavenumber, indicating they arise from linear perturbation dynamics.
Moreover, they appear at precisely the wavenumber predicted by (linear) Floquet theory.
This provides very strong evidence that the effects we are seeing are a well understood physical effect associated with evolution by the coupled GPEs.

\subsection{Effective Lattice Wavenumber for Discrete Laplacian Stencils}
In this subsection we briefly derive the linear dispersion relationship associated with a discrete approximation to the Laplacian operator.
The distortion of the effective wavenumber for linear fluctuations from the continuum limit by the use of finite differencing stencil is well-known, but is presented here for completeness.

\begin{figure}
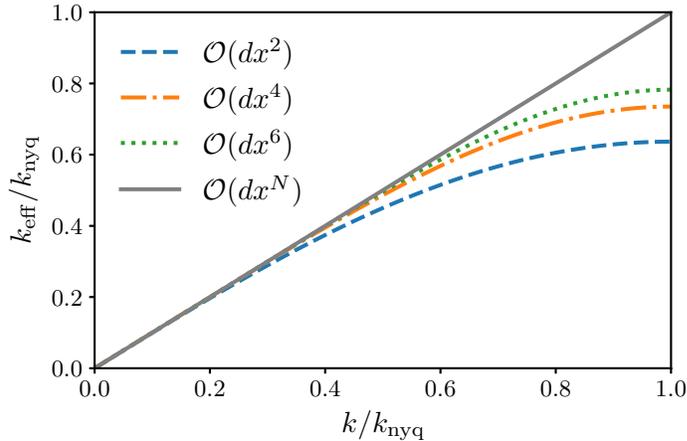
\centering
  \includegraphics[width=\singleWid]{{{effective-k}}}
  \caption{Effective wavenumber (defined in~\eqref{eqn:keff}) for a second (\emph{blue dashed}), fourth (\emph{orange dot-dashed}), and sixth (\emph{green dotted}) order accurate symmetric discrete Laplacian stencil as a function of the true wavenumber in one-dimension.  For comparison, we also plot the relationship for the pseudospectral stencil (\emph{gray solid}) up to the Nyquist frequency (labelled as $\mathcal{O}(dx^N)$), which matches the continuum value for all wavenumbers below the Nyquist.}
  \label{fig:effective-k}
\end{figure}

Consider a $d$-dimensional discrete lattice with sites $\xLat{\vi}$ labelled by the vector index $\vi = (m_1,m_2,\dots,m_d)$.
We will denote function values at position $\xLat{\vi}$ by $f_{\vi} \equiv f(\xLat{\vi})$.
For simplicity, assume that the lattice sites are arranged in a rectangular grid with uniform lattice spacing $dx$, so that $\xLat{\vi} = \vi dx$.
Consider finite-difference stencils for the Laplacian operator of the form
\begin{equation}
  \nabla^2 f(\xLat{\vi}) \approx L^{{\rm (D)}}[f](\xLat{\vi}) \equiv \frac{1}{dx^2}\sum_{\vj} \cSten{\vj}\left[f(\bm{x}_{\vi+\vj}) - f(\xLat{\vi})\right]
\end{equation}
with the vector indices $\vj \in \{(\alpha_1,\alpha_2,\dots,\alpha_d), \alpha_i=-\frac{N}{2},-\frac{N}{2}+1,\dots,\frac{N}{2}-1 \}$.  The choice of coefficients $\cSten{\vj}$ specify the stencil.
Now consider a linear operator $\op{\mathcal{O}}$ of the form
\begin{equation}
  \label{eqn:linear-operator}
  \op{\mathcal{O}}[f] = -L^{{\rm (D)}}[f] + \mathcal{C}f \approx \left(-\nabla^2+\mathcal{C}\right)f
\end{equation}
with $\mathcal{C}$ a constant.
For simplicity, we consider only a single function $f$ defined on our lattice.  
The generalization to many linearly coupled fields is straightforward.
Taking a discrete Fourier transform of~\eqref{eqn:linear-operator}, we obtain in one-dimension
\begin{equation}
  \label{eqn:ft-lin-op}
  \tilde{\mathcal{O}}[f](\kVec) \equiv \sum_{\vi} e^{i\kVec\cdot\xLat{\vi}}\left(-L_i^{{\rm (D)}}[f] + \mathcal{C}f_i\right) = \left(\frac{1}{dx^2}\sum_{\vj}\cSten{\vj}\left[1-e^{i\kVec\cdot\vj dx}\right] + \mathcal{C}\right)\tilde{f}_k
\end{equation}
with
\begin{equation}
  \tilde{f}_{\kVec} \equiv \sum_{\vi} e^{i\kVec\cdot\xLat{\vi}}f(\xLat{\vi}) \, .
\end{equation}
In the continuum, the RHS of~\eqref{eqn:ft-lin-op} would be $(k^2 + C)\tilde{f}_k$, so we identify an effective wavenumber $k_{\rm eff}(\kVec)$ associated with the Laplacian stencil as a function of the continuum wavenumber $\kVec$
\begin{align}
  k_{\rm eff}^2(\kVec) &= \frac{1}{dx^2}\sum_{\vj}\cSten{\vj}\left(1-e^{i\kDa{\vj} dx}\right) \notag \\
                      &= \frac{1}{dx^2}\sum_{\vj}\left[\frac{\cSten{\vj}+\cSten{-\vj}}{2}\left(1-\cos\left(\pi\frac{\kDa{\vj}}{\kNyq}\right)\right) - i\frac{\cSten{\vj}-\cSten{-\vj}}{2}\sin\left(\pi\frac{\kDa{\vj}}{\kNyq}\right)\right] \, ,
\end{align}
where $\kNyq = \frac{\pi}{dx}$ is the Nyquist wavenumber.
To avoid spurious numerical damping associated with asymmetric stencils, we use a symmetric stencil $c_{\vj} = c_{-\vj}$ so that the imaginary contribution vanishes, and we have
\begin{equation}\label{eqn:keff}
  \frac{k_{\rm eff}^2(\kVec)}{\kNyq^2} = \frac{1}{\pi^2}\sum_{\vj} c_{\vj}\left[1-\cos\left(\pi\frac{\kDa{\vj}}{\kNyq}\right)\right] \, .
\end{equation}
In~\figref{fig:effective-k}, we show this effective wavenumber for several low-order symmetric one-dimensional Laplacian stencils.

We also note one further complication to the picture above.
Our code evolves the condensate fields $\wf_i$, and the numerical Laplacian is thus defined to act directly on these fields rather than the densities $\dens_i$ and phases $\phase_i$.
However, our analytic derivation of the Floquet instability was done in the density and phase variables.
Because of the nonlinear transformation between these sets of variables, additional distortions to the required numerical stencils appear in the equations of motion for $\rho_i$ and $\phi_i$.
However, these distortions vanish in the limit of strictly linear inhomogeneities.
The excellent match between our analytic predictions for the location of the Floquet band and the finite-differencing results is thus even further evidence that the new physics seen in our nonlinear simulations is simply the manifestation of the linear Floquet instability.

\section{Numerical Convergence and Conservation Tests}\label{app:convergence}
In this appendix we present a variety of convergence and consistency tests to illustrate the precision of our numerical approach.
We provide two tests, including: a direct test of numerical convergence with variation of either the time step $dt$ or grid spacing $dx$; and a demonstration that relevant conserved charges are time-invariant.
These tests will demonstrate that we are correctly solving for the condensate dynamics under the assumption they are described by the coupled GPEs.
Of course, more precise physical modeling will result in corrections to the GPE description.
However, these corrections are \emph{not} modeled by the errors introduced by approximate numerical methods.

Throughout this appendix, we consider convergence properties for two fiducial simulations presented in the main text: the simulation in the top left panel of~\figref{fig:stabilise-vary-delta} (with $\lPar = 0$ and no Floquet modes), and the simulation shown in~\figref{fig:floquet-nonlinear} (with $\lPar = 1.3$ and Floquet modes).
For reference, the parameters are
\begin{equation}
  \frac{\nuZero}{\gS\rhoBG} = 2\times 10^{-3}\, , \qquad \lPar = 0\, ,
\end{equation}
and
\begin{equation}
  \frac{\nuZero}{\gS\rhoBG} = 2\times 10^{-3}\, , \qquad \lPar = 1.3\, , \qquad \frac{\nuFreq}{\gS\rhoBG} = 50\times 2\sqrt{\frac{\nuZero}{\gS\rhoBG}} \approx 4.47 \, ,
\end{equation}
respectively.
The initial conditions for each simulation (including the particular realization of the fluctuations) were the same.
The former simulation doesn't possess a false vacuum, and thus experiences a spinodal instability as shown in~\secref{sec:nonlinear-scalar} rather than nucleating bubbles.
However, by removing the explicit time-dependence, the energy of the system will be conserved and thus provide us an additional diagnostic tool for testing our numerics.
As well, the Floquet instabilities will not be present in this case, allowing spatial convergence to be tested in a much simpler way.
In order to allow a direct comparison between simulations with different time steps or grid spacings, the initial conditions are identical for each simulation.
There are several dynamical timescales that must be resolved by our choice of timestep $dt$.
In the linear regime, these are roughly the driving frequency $\omega$, the linear frequency of the Nyquist mode for the total ($\omega_{\rm tot}$) and relative ($\omega_{\rm rel}$) phonons, and the growth rate of the Floquet modes.\footnote{Technically, the oscillation frequencies should be obtained from the imaginary parts of the Floquet exponents, but $\omega_{\rm tot}$ and $\omega_{\rm rel}$ can be used as reasonable proxies.}
The free oscillation frequencies of the Nyquist modes can be easily obtained (see Braden \etal~\cite{Braden:2017add})
\begin{subequations}
\begin{align}
  \frac{\hbar\omega_{\rm rel}}{\gS \rhoBG} &= 4\sqrt{\frac{\hbar^2k_{\rm nyq}^2}{4\gS\rhoBG m} - \frac{\nuZero}{\gS\rhoBG}}\sqrt{1 + \frac{\hbar^2k_{\rm nyq}^2}{4\gS\rhoBG m} - \frac{\nuZero}{\gS\rhoBG}} \\
  \frac{\hbar\omega_{\rm tot}}{\gS \rhoBG} &= \frac{\hbar k_{\rm nyq}}{\sqrt{\gS\rhoBG m}}\sqrt{1+ \frac{\hbar^2k_{\rm nyq}^2}{4\gS\rhoBG m}} \, .
\end{align}
\end{subequations}
Since $\frac{\nuZero}{\gS\rhoBG} \ll 1$, we have $\omega_{\rm rel} \approx \omega_{\rm tot}$.
To obtain spatial resolution, we will be interested in the limit where the Floquet band is below the Nyquist mode, which occurs roughly when $\nuFreq = 2\omega_{\rm rel}$.
Therefore, for spatially resolved simulations, we expect that $\omega_{\rm tot}$ and $\omega_{\rm rel}$ will dominate the temporal convergence properties.
Of course, once the fields become strongly nonlinear, new timescales may emerge, and the interactions between different modes may modify the effective oscillation frequencies much like plasma effects.  However, we will take the linear scales given above as a guideline.

\subsection{Direct convergence tests}
First we show direct pointwise convergence tests with respect to changes in both the time step $dt$ and grid spacing $dx$.
These demonstrate the rapid convergence displayed by our Gauss-Legendre time stepping and pseudospectral discretization, respectively.
To explore pointwise convergence, we consider the following norm between two simulations $\wf^{(n)}$ and $\wf^{(n+1)}$
\begin{equation}\label{eqn:max-norm}
  \abs{\wf^{(n+1)}-\wf^{(n)}}_{\rm max} \equiv \max{\left\{\abs{\wf^{(n+1)}(x_i)-\wf^{(n)}(x_i)} : x_i \in \mathbb{L}\right\}} \, ,
\end{equation}
which measures the maximal pointwise difference between the two simulations over the entire simulation volume.
Here, the superscript ${}^{(n)}$ labels the simulations, and $\mathbb{L}$ is a collection of spatial points at which to evaluate our functions.
In our convergence testing, we will consider differences between simulations in which either the time step $dt$ or grid spacing $dx$ differ by a factor of $2$, while holding the grid spacing or time step fixed respectively.
When only the time step is varied, the spatial grids match exactly, and we simply have to ensure that we compare simulations at the same time.
For our equally spaced Fourier collocation grid, each subsequent spatial grid refinement is a superset of the previous one, so the comparison can be done directly on the coarser grid.
Although we expect to see rapid convergence because of our use of highly accurate numerical methods, we also expect the presence of exponentially growing Floquet modes to slowly degrade the quality of the convergence for two reasons.
First, our temporal integrator is symplectic, and thus preserves phase space volumes.
For linear fluctuations, this corresponds to accurately tracking the overall amplitude of oscillating Fourier modes.
However, the numerical tradeoff for this is a small error in the oscillation phase, and a corresponding error in pointwise comparisons at a fixed time.
For exponentially growing modes, this error will grow with time as the modes grow in amplitude.
For highly nonlinear fluctuations, there will generally be localized stuctures that will propagate through the simulation volume.
Small temporal phase errors lead to small changes in the shapes and velocities of these structures.
Over time, the same structure present in two simulations with different choices of $dt$ may follow a slightly different spacetime path, leading to a growing pointwise difference between the simulations.

\begin{figure}
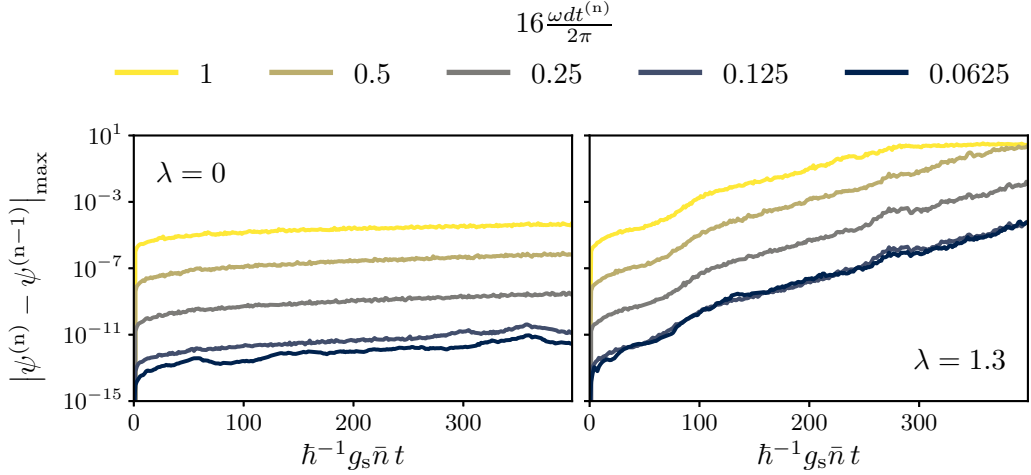

  \includegraphics[width=\fullWid]{{{gpe-convergence-dt}}}
  \caption{Pointwise convergence of our numerical code as the numerical time step $dt$ is varied.  The colors run from yellow to blue as the time step $dt$ is decreased.  \emph{Left}: Convergence properties for a system undergoing spinodal instability with Floquet modes induced by external coupling modulation. \emph{Right}: Convergence for a proposed false vacuum simulation including the effects of exponentially growing Floquet modes.  In each case, the $\mathcal{O}\left(dt^{10}\right)$ convergence is clearly present for the longer choices time steps $dt$.  For smaller values of $dt$, however, we instead see a saturation due the effects of machine precision roundoff errors.  In each case, we have $\frac{\sqrt{\gS\rhoBG m}}{\hbar}dx = \frac{50}{1024}\sqrt{\frac{\gS\rhoBG}{4\nuZero}}$ and $\frac{\nuFreq dt}{2\pi} = \frac{1}{2^n}$ for $n=4,5,6,7,8$, where $n$ refers to the smaller of the two time steps in the comparison between simulations.}
  \label{fig:convergence-test-dt}
\end{figure}

\begin{figure}
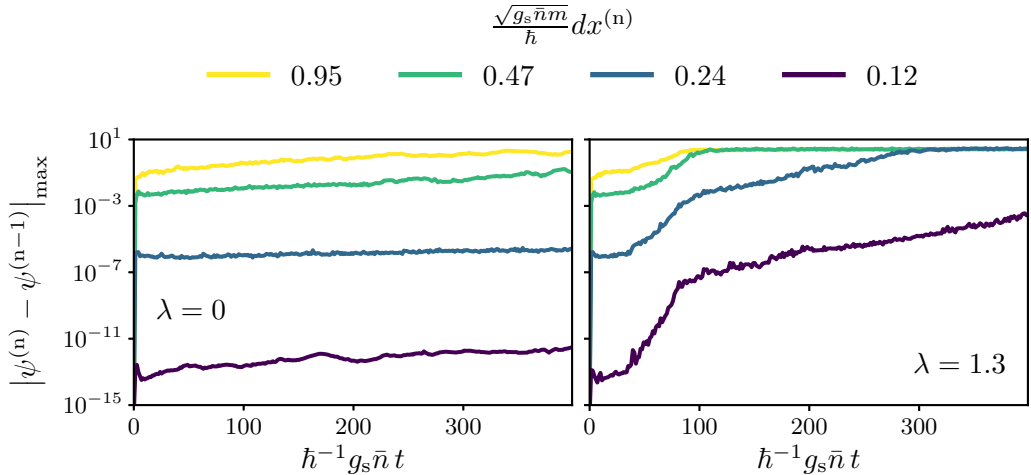

  \includegraphics[width=\fullWid]{{{gpe-convergence-dx}}}
  \caption{Pointwise convergence of our numerical code as the numerical grid spacing $dx$ is varied for a simulation of a second-order phase transition without Floquet modes (\emph{left}) and the proposed false vacuum decay setup with Floquet modes (\emph{right}).  The colors run from yellow to purple as the grid spacing $dx$ is decreased.  In both cases, we see an extremely rapid convergence to an eventual saturation level as the grid spacing is decreased.  However, while the saturation level is constant in the absence of the growing Floquet modes, it increases with time when the Floquet modes are excited.  This matches the convergence with the timestep shown in~\figref{fig:convergence-test-dt}.  As explained in the main text, this is most likely due to the exponential growth of the Floquet modes combined with slight numerical errors in the oscillation phase.  We have taken ${\frac{\sqrt{\gS\rhoBG m}}{\hbar}dx = \frac{50}{295}\sqrt{\frac{\gS\rhoBG}{4\nuZero}}\frac{1}{2^n}}$ for $n=1,2,3,4$.  For each choice of $dx$, the time-step was taken to ensure saturation of the temporal convergence.}
  \label{fig:convergence-test-dx}
\end{figure}

Below we show the maximal pointwise norm~\eqref{eqn:max-norm} between simulation pairs with either $dt$ (\figref{fig:convergence-test-dt}) or $dx$ (\figref{fig:convergence-test-dx}) varied.
As the time step is decreased, initially we see the pointwise error decrease by a factor of $2^{10}\sim 1000$, as expected for a tenth-order accurate integrator.
This eventually saturates at a level of $\mathcal{O}(10^{-13})$ for the simulations without Floquet modes present, while the saturation level increases with time in the presence of Floquet modes.
As explained above, this is not unexpected due to the exponential growth of the Fourier modes, and the tradeoff made in temporal phase and amplitude accuracy described above.
To further identify the exponentially growing Floquet modes as the root cause of the time-dependent saturation of the temporal accuracy, we also performed the same temporal convergence tests for ${\frac{\sqrt{\gS\rhoBG m}}{\hbar}dx = 1.89}$ and ${\frac{\sqrt{\gS\rhoBG m}}{\hbar}dx = 1.79}$ for the ${\lPar=1.3}$ simulation, corresponding to a grid-spacing that just excludes and just includes the unstable Floquet band.
We found a growing saturation level for the latter case, while the convergence plateaued in a manner similar to the ${\lPar = 0}$ simulation for the former case.
As a test that these errors did not arise from aliasing into long-wavelength fluctuations, we performed the corresponding convergence tests with a finite-differencing Laplacian stencil (which does not lead to aliasing) and found the same growth in the saturation level.

Meanwhile, as the spatial grid spacing is decreased we see an exponential improvement in the pointwise convergence owing to our use of a pseudospectral scheme.
In particular, for the simulations shown here, an increase in the number of grid points by a factor of $8$ leads to over a ten order of magnitude improvement in convergence without the chaotic behaviour induced by the Floquet band.
As with the temporal convergence, we see the saturation level increase with time for the simulation including Floquet modes.

\subsection{Conserved charge preservation}
The previous subsection explicitly demonstrated the rapid convergence properties of our temporal and spatial discretization schemes.
We also observed a saturation of the pointwise accuracy of our simulations, which we believe arises from the exponential linear growth of fluctuation amplitudes and the resulting chaotic dynamics in the full field space.
Since we lack an exact analytic solution to compare to, technically these results demonstrate that we are converging to a solution, but not necessarily the correct solution of the equations of motion.
To demonstrate that we are indeed solving the coupled GPEs, we now show that our numerical scheme preserves the various conserved quantities of the continuum equations.
As emphasized above, the goal of this work is to explicitly demonstrate the need for modeling of the short wavelength behavior of the condensate, not to explore how these corrections modify the false vacuum decay picture.
Therefore, we work under the assumption that the coupled GPEs are indeed the correct evolution equations.
Of course, the various phenomenological corrections briefly mentioned in~\secref{sec:floquet-modes} will lead to a violation of these conserved charges, but this will be the subject of future work.

\begin{figure}
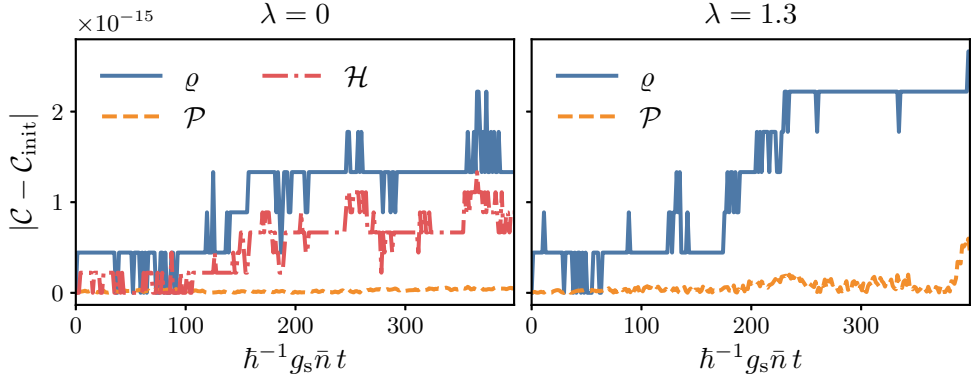

  \includegraphics[width=\fullWid]{{{noether-charges}}}
  \caption{\emph{Left}: Time evolution of the conserved Noether charges: the total density $\densTot$ (\emph{blue solid}), momentum $\mathcal{P}$ (\emph{orange dashed}), and energy $\mathcal{H}$ (\emph{red dot-dashed}).  For simplicity, we collectively denote them by $\mathcal{C}$.   We have also removed the initial spatial average $\mathcal{C}_{\rm init}$ to isolate the temporal variations.  We consider both a time-independent (\emph{left}) and time-dependent (\emph{right}) $\nuT$.  In the former case all of the quantities are conserved by the equations of motion, while in the latter the energy density (\ie Hamiltonian) is no longer constant due to the explicit time-dependence of the Hamiltonian.  In both cases, we see that our numerical scheme leads to machine-precision conservation of all conserved charges.  This is especially evident for the number density, and total energy density, where discrete steps from machine roundoff are clearly visible.}
  \label{fig:conserved-charges}
\end{figure}

When $\nuAmp = 0$, we have the following three locally conserved charges
\begin{enumerate}
  \item the total condensate particle density $\densTot \equiv \sum_i\abs{\wf_i}^2$, 
  \item the total momentum density $\mathcal{P} \equiv \frac{i}{2}\sum_i\left(\nabla\wf_i^\dagger \wf_i - \wf_i^\dagger\nabla\wf_i \right) = \sum_i\dens_i\nabla\phase_i$, and
  \item the condensate energy density $\mathcal{H} \equiv \sum_i\left(\frac{\hbar^2}{2m}\abs{\nabla\wf_i}^2 + \frac{g}{2}\abs{\wf_i}^4\right) - \nu\left(\wf_1^\dagger\wf_2+\wf_1\wf_2^\dagger\right)$.
\end{enumerate}
These are readily identified, respectively, with invariance of the action under a global phase rotation of the $\wf_i$'s, spatial translations, and time translation.
When $\nuAmp \neq 0$ we lose time translation invariance, and the energy density is no longer conserved.
Although not considered here, if we further consider inhomogeneous condensates evolving in an external potential, then the spatial translation invariance will be broken and momentum no longer conserved.  However, even in this limit the total particle number will be conserved, which is the origin of the ungapped linear fluctuations associated with the total phase phonons.\footnote{In the limit $\nuT = 0$, our coupled GPEs decouple into two independent condensates, each satisfying the nonlinear Schrodinger equation (NLS).  Therefore, in this limit the individual condensate densities, momenta, and energies are conserved. Further, in $1$+$1$-dimensions the NLS is integrable, with an infinite hierarchy of additional conserved quantities.  We will not consider this limit here, although how the introduction of $\nuT$ leads to the breaking of these conserved quantities is interesting.}
In~\figref{fig:conserved-charges}, we demonstrate the preservation of these conserved quantities for our fiducial convergence testing simulations.
To make a fair comparison, we choose ${\frac{\sqrt{\gS\rhoBG m}}{\hbar}dx = \frac{50\times 2}{4720}\frac{\nuZero}{\gS\rhoBG} \approx 0.19}$ and ${\frac{\gS\rhoBG}{\hbar} dt = \frac{2\pi}{256}\frac{\gS\rhoBG}{\hbar\nuFreq} \approx 5.49\times 10^{-3}}$ for both simulations.
In all cases, we obtain excellent preservation of all of the Noether charges for the simulations both with and without the Floquet instability.
The results are particularly striking for the mean total density, where discrete jumps are visible associated with the finite precision of machine arithmetic.
Similar discrete jumps also occur for the energy density for the case where the Hamiltonian is time-independent.
When $\nuT$ oscillates in time and the Floquet modes are present in the system, the external driver $\nuT$ drives exponential growth of the unstable modes.
As a result, energy is injected into the system, leading to a growth in the mean energy density with time.
We verified that this is the case, but do not include the results in this appendix as here we want to focus on only the conserved charges.

For simplicity, here we have shown the conservation properties for a single choice of spatial and temporal resolution.
However, the accuracy of the conservation laws is sensitive to the choice of $dx$ and $dt$.
Although not presented here, we also investigated how well the conserved charges were preserved for the same suite of simulations used in the direct convergence testing shown in~\figref{fig:convergence-test-dt} and~\figref{fig:convergence-test-dx}.
From these investigations, we found that the preservation of the total density $\densTot$ was primarily sensitive to the choice of time-step $dt$, with little sensitivity to the choice of grid-spacing $dx$.
Meanwhile, the preservation of the total momentum $\mathcal{P}$ had a saturation level with decreasing $dt$ that was sensitive to the choice of lattice spacing.
Finally, the energy density $\mathcal{H}$ displays convergence properties intermediate between the total density and momentum, with more sensitivity to the grid-spacing $dx$ than the number density, but spatial saturation occuring at a larger grid spacing than for the momentum.

\end{document}

%% file: notation.tex
\newcommand{\abs}[1]{\ensuremath{\left|#1\right|}}
\newcommand{\op}[1]{\ensuremath{\hat{#1}}}

\newcommand{\ud}[2]{\ensuremath{\mathrm{d}^{#2}#1\,}}
\newcommand{\dd}{\ensuremath{\mathrm{d}}}

\newcommand{\aveV}[1]{\ensuremath{\langle #1 \rangle_{\rm V}}}

\newcommand{\bg}[1]{\ensuremath{\bar{#1}}}
\newcommand{\pert}[1]{\ensuremath{\delta#1}}
\newcommand{\nodim}[1]{\ensuremath{\bar{#1}}}

\newcommand{\wf}{\ensuremath{\psi}}

\newcommand{\phase}{\ensuremath{\phi}}
\newcommand{\dens}{\ensuremath{\rho}}

\newcommand{\phaseTot}{\ensuremath{\vartheta}}
\newcommand{\phaseRel}{\ensuremath{\varphi}}
\newcommand{\densTot}{\ensuremath{\varrho}}
\newcommand{\densRel}{\ensuremath{\varepsilon}}

\newcommand{\tScl}{\ensuremath{\omega_{\rm p}}}
\newcommand{\xScl}{\ensuremath{\kappa_{\rm p}}}
\newcommand{\rhoScl}{\ensuremath{\Lambda}}
\newcommand{\csNum}{\ensuremath{c_{\rm p}}}
\newcommand{\enScl}{\ensuremath{E_{\rm p}}}

\newcommand{\lPar}{\ensuremath{\lambda}}
\newcommand{\gS}{\ensuremath{g_{\rm s}}}
\newcommand{\gC}{\ensuremath{g_{\rm c}}}
\newcommand{\gD}{\ensuremath{\delta g}}
\newcommand{\gSScl}{\ensuremath{\nodim{g}_{\rm s}}}
\newcommand{\gCScl}{\ensuremath{\nodim{g}_{\rm c}}}
\newcommand{\gDScl}{\ensuremath{\delta \nodim{g}}}

\newcommand{\nuT}{\ensuremath{\nu}}
\newcommand{\nuZero}{\ensuremath{\nu_0}}
\newcommand{\nuAmp}{\ensuremath{\delta}}
\newcommand{\nuFreq}{\ensuremath{\omega}}

\newcommand{\rhoBG}{\ensuremath{\bar{n}}}

\newcommand{\vi}{\ensuremath{\vec{{\rm m}}}}
\newcommand{\vj}{\ensuremath{\bm{\alpha}}}

\newcommand{\cSten}[1]{\ensuremath{c_{#1}}}
\newcommand{\kVec}{\ensuremath{\bm{k}}}
\newcommand{\xLat}[1]{\ensuremath{\bm{x}_{#1}}} 
\newcommand{\kNyq}{\ensuremath{k_{\rm nyq}}}
\newcommand{\kDa}[1]{\kVec\cdot\vj}

\newcommand{\cphi}{\ensuremath{\sigma}}